\documentclass[12pt, leqno]{article}
\usepackage{amsmath,setspace,multirow,lineno}
\usepackage[font=small,labelfont=bf,singlelinecheck=off]{caption}
\usepackage[top=1.42in, bottom=1.42in, left=1.1in, right=1.1in]{geometry}

\usepackage[round]{natbib}
\usepackage{color,soul}

\DeclareCaptionStyle{italic}[justification=centering]{labelfont={bf},textfont={it},labelsep=colon}

\usepackage{graphicx,psfrag,epsf,textcomp,epstopdf, amsthm, paralist}
\usepackage{enumerate, dsfont, alltt, verbatim, algorithm, algpseudocode, algorithmicx}
\usepackage{url,xcolor}
\usepackage{booktabs, subfig, bm, tabularx, paralist,mathpazo,tikz,longtable,microtype, authblk}
\usepackage[utf8]{inputenc}

\usepackage[pdftex,colorlinks=true]{hyperref}
\definecolor{darkblue}{rgb}{0,0,.6}
\hypersetup{citecolor=darkblue,linkcolor=darkblue,urlcolor=darkblue}
\definecolor{DarkRed}{rgb}{.7,0,.4}

\usepackage{comment,orcidlink}

\newcommand{\blind}{0}

\newcommand{\X}{\mathcal{X}}

\newcommand{\Rlogo}{\protect\includegraphics[height=1.8ex,keepaspectratio]{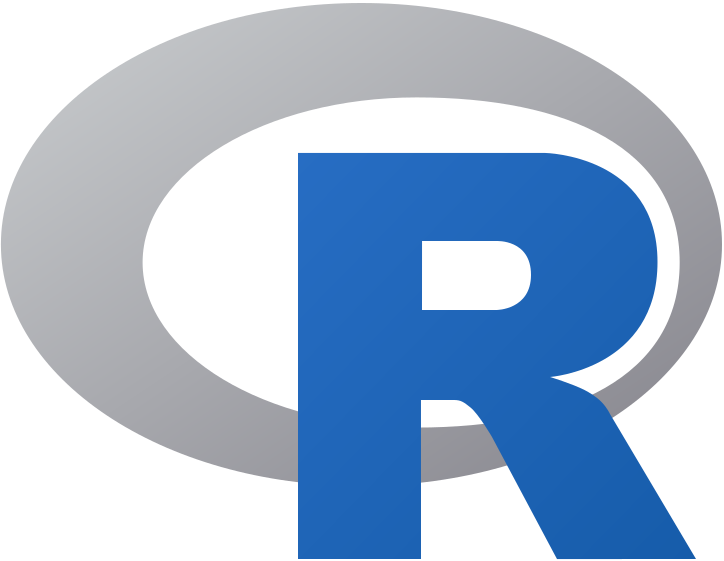}}

\graphicspath{{plots/}}

\newsavebox\CBox

\newtheorem{@definition}{\sc Definition}[section]

\renewcommand\X{\mathcal{X}}

\date{}

\begin{document}

\def\spacingset#1{\renewcommand{\baselinestretch}{#1}\small\normalsize} \spacingset{1}

\if0\blind
{
\title{\bf Spatial Functional Deep Neural Network Model: \mbox{A New Prediction Algorithm}}}
\author[1]{\normalsize Merve Basaran}
\author[1]{\normalsize Ufuk Beyaztas\thanks{Corresponding address: Department of Statistics, Marmara University, 34722, Kadikoy-Istanbul, Turkey; Email: ufuk.beyaztas@marmara.edu.tr} \orcidlink{0000-0002-5208-4950}}
\author[2]{\normalsize Han Lin Shang \orcidlink{0000-0003-1769-6430}}
\author[3]{\normalsize Zaher Mundher Yaseen \orcidlink{0000-0003-3647-7137}}

\affil[1]{\normalsize Department of Statistics, Marmara University, Turkey}
\affil[2]{\normalsize Department of Actuarial Studies and Business Analytics, Macquarie University, Australia}
\affil[3]{\normalsize Civil and Environmental Engineering Department, King Fahd University of Petroleum \& Minerals, Saudi Arabia}

\maketitle
\fi

\if1\blind
{
\title{\bf Spatial Functional Deep Neural Network}
\author{}
} \fi

\maketitle

\begin{abstract}

Accurate prediction of spatially dependent functional data is critical for various engineering and scientific applications. In this study, a spatial functional deep neural network model was developed with a novel non-linear modeling framework that seamlessly integrates spatial dependencies and functional predictors using deep learning techniques. The proposed model extends classical scalar-on-function regression by incorporating a spatial autoregressive component while leveraging functional deep neural networks to capture complex non-linear relationships. To ensure a robust estimation, the methodology employs an adaptive estimation approach, where the spatial dependence parameter was first inferred via maximum likelihood estimation, followed by non-linear functional regression using deep learning. The effectiveness of the proposed model was evaluated through extensive Monte Carlo simulations and an application to Brazilian COVID-19 data, where the goal was to predict the average daily number of deaths. Comparative analysis with maximum likelihood-based spatial functional linear regression and functional deep neural network models demonstrates that the proposed algorithm significantly improves predictive performance. The results for the Brazilian COVID-19 data showed that while all models achieved similar mean squared error values over the training modeling phase, the proposed model achieved the lowest mean squared prediction error in the testing phase, indicating superior generalization ability. 

\end{abstract}


\newpage
\spacingset{1.65} 

\section{Introduction} \label{sec1}

Recent technological advancements have revolutionized the way data sets are collected, leading to increasingly sophisticated tools capable of capturing complex (non)linear structures \cite{hong2008model, khemani2024review}. In a graphical representation of curves, images, shapes, and more generally, objects residing in infinite-dimensional spaces, functional data set has become a fundamental component in numerous scientific domains, including energy, geosciences, finance, agriculture, and healthcare \cite{zhang2004review}. The growing prevalence of such data has necessitated the development of advanced statistical methodologies tailored to functional data analysis \cite{storey2017big}. This field has witnessed substantial progress, particularly in the formulation and refinement of regression models that establish relationships between functional and scalar variables. For a comprehensive overview of theoretical advancements and recent methodological developments in functional data analysis, we refer the reader to \cite{Ramsay1991}, \cite{Ramsay2006}, \cite{Horvath2012}, and \cite{Kokoszka2017}.

Among others, scalar-on-function regression (SoFRM) has gained prominence, as it effectively links functional predictors to scalar responses, thereby enriching statistical modeling frameworks \citep[see, e.g.,][]{Hastie1993, Ramsay2006, Morris2015, Reiss2017}. The recent integration of neural networks into functional data analysis has expanded the methodological landscape, moving beyond classical linear approaches and enabling more flexible and robust modeling strategies \citep[see, e.g.,][]{WC23, Thind2023}.

Let us consider a random sample $[Y_i, \X_i(u)]$, $i \in \{1, \ldots, n\}$ drawn from the joint distribution of $(Y, \X)$, where the response variable $Y \in \mathbb{R}$ is scalar, and the predictor $\X = \{\X(u)\}_{u \in \mathcal{I}}$ is a stochastic process with finite second-moment exhibiting complex spatial dependence. The trajectories of $\X$ are assumed to belong to the Hilbert space $\mathcal{L}^2(\mathcal{I})$, where $\mathcal{I}$ represents a compact and bounded subset of~$\mathbb{R}$. The SoFRM is then formulated as:
\begin{equation} \label{eq:sof}
Y = \beta_0 + \int_{\mathcal{I}} \X(u) \beta(u) \, du + \epsilon,
\end{equation}
where $\beta_0 \in \mathbb{R}$ is the intercept, $\beta(u) \in \mathcal{L}^2(\mathcal{I})$ denotes the regression coefficient function, and $\epsilon$ represents the random error term, which is assumed to be independent and identically distributed, typically following a normal distribution with mean zero and variance $\sigma^2$.

The linear SoFRM given in~\eqref{eq:sof} has been widely extended in the literature to accommodate more flexible modeling approaches, including nonlinear and nonparametric frameworks \citep[see, e.g.,][]{James2005, Yao2010, Muller2013, Reiss2017}. Despite these advancements, a fundamental assumption underlying many of these models is the independence of observations within the sample. However, in many real-world applications, such as energy, geosciences, finance, agriculture, and healthcare, data often exhibit spatial correlations that violate this assumption. As a result, it is crucial to develop analytical methodologies that explicitly incorporate spatial dependence to ensure more accurate and reliable statistical inference.

Various statistical models have been introduced to spatially correlated functional data to capture spatial dependencies, including the spatial autoregressive model, the spatial error model, and the spatial Durbin model \citep[see, e.g.,][]{Bouka2023, Hu2021, Huang2021, Pineda2019, Medina2011, Medina2012, Hu2020, BSM2025}. Among these, the spatial autoregressive model is particularly significant, as it incorporates spatial lag in the response variable through a single spatial dependence parameter, effectively accounting for interactions across spatial units. Motivated by this, we consider the spatial scalar-on-function regression model (SSoFRM) to investigate spatial dependencies in functional datasets across regions. The model is formulated as:
\begin{align}
Y &= \beta_0 \bm{1}_n + \rho \bm{W} Y + \int_{u\in \mathcal{I}} \X(u) \beta(u) \, du + \epsilon, \label{eq:ssofr} \\
&= (\mathbb{I}_n - \rho \bm{W})^{-1} \beta_0 \mathbf{1}_n + (\mathbb{I}_n - \rho \bm{W})^{-1} \int_{u\in \mathcal{I}} \X(u) \beta(u) \, du + (\mathbb{I}_n - \rho \bm{W})^{-1} \epsilon, \nonumber
\end{align}
where $\mathbb{I}_n$ is an $n \times n$ diagonal matrix, $\bm{1}_n$ is an $n$-dimensional vector of ones, $\rho \in (-1, 1)$ represents the spatial autocorrelation parameter,  $\bm{W} = (w_{i i^\prime})_{1 \leq i, i^\prime \leq n}$ is an $n \times n$ spatial weight matrix, where each element $w_{i i^\prime}$ quantifies the spatial proximity or connectivity between locations $i$ and $i^\prime$, and $\mathcal{I}$ denotes a function support, commonly a unit interval. 

A comprehensive review of existing SSoFRM models highlights a predominant reliance on linear structures to capture the relationship between the response and explanatory variables \cite{reiss2017methods}. While these linear formulations can yield accurate predictions, real-world phenomena often exhibit intricate nonlinear dependencies and complex interaction effects, which conventional models fail to address adequately \cite{lin2022linear}. In addition, existing SSoFRMs are limited to a single functional predictor, restricting their applicability in scenarios where multiple functional and scalar covariates are necessary for modeling spatially dependent scalar responses. 

To overcome these limitations, the current research introduced a nonlinear spatial SSoFRM (NSSoFRM) that integrates both \textit{nonlinear interactions} and \textit{multiple covariates} within the SSoFRM framework. The proposed model allows for the inclusion of multiple functional and scalar predictors, making it more adaptable to real-world applications where mixed-type covariates are essential for capturing spatially structured dependencies. For parameter estimation, inspired by the published work by \cite{Thind2023}, deep neural networks was extended to accommodate functional observations, leading to a functional deep neural network (FDNN) that serves as an automated predictive framework.

A natural nonlinear extension of~\eqref{eq:ssofr} involves introducing a link function $g(\cdot)$ applied to the entire linear predictor, including the spatially lagged response term. However, such an approach is generally problematic in spatial modeling: Suppose the spatial dependence term $\rho \bm{W} Y$ undergoes a nonlinear transformation prior to estimating $Y$, the inherent dependence structure of the spatial process is disrupted, leading to model misspecification. This issue arises because spatial autoregressive models fundamentally rely on the inverse mapping $(\mathbb{I}_n - \rho \bm{W})^{-1}$ to propagate spatial effects across neighboring units in a well-defined manner. Introducing $g(\cdot)$ before applying this inverse transformation alters the equilibrium structure governing spatial interactions, which can result in inconsistent estimation of $\rho$ and a misrepresentation of how neighboring observations influence the response variable. 

To address these challenges, we employ a hybrid estimation approach that first isolates the spatial dependence structure before incorporating nonlinear interactions. We begin by estimating the spatial autoregressive parameter $\rho$, denoted as $\widehat{\rho}$, using the maximum likelihood (ML) estimator to ensure proper inference of spatial dependence. Once $\widehat{\rho}$ is obtained, we apply the inverse transformation $(\mathbb{I}_n - \rho \bm{W})^{-1}$ to the linear predictor, thereby integrating spatial dependence into the covariate structure while maintaining the correct dependence assumptions. The transformed predictor is then used as input to the FDNN, where a nonlinear mapping $g(\cdot)$ captures complex relationships between the spatially adjusted functional and scalar covariates and the expected response. This methodology retains the spatial dependence framework while allowing for flexible nonlinear interactions between predictors and the response variable.

In the second step of our approach, the Adam optimization algorithm is utilized to determine the optimal functional and scalar weight parameters, with several hyperparameters, such as the number of layers, neurons per layer, decay rate, and early stopping, carefully tuned to enhance predictive accuracy. By incorporating nonlinear interactions, multiple functional and scalar covariates, and spatial dependencies into a unified modeling framework, the proposed NSSoFRM enhances the effectiveness of spatial functional regression, surpassing the capabilities of several existing approaches.

Our proposed SFDNN framework aligns with a growing body of work that integrates deep learning with functional and spatial data analysis. Recent studies have demonstrated the potential of FDNNs in classification and regression with multidimensional functional inputs, as seen in \cite{Wang2023FDNN} who establish minimax optimality for FDNN classifiers applied to non-Gaussian multidimensional functional data. Furthermore, hybrid modeling strategies that combine deep neural architectures with spatial or spatiotemporal structures have gained traction in the statistics community. The review by \cite{Wikle2023Review} comprehensively outlines approaches such as deep Gaussian processes, deep hierarchical models, and convolutional/recurrent neural networks adapted for spatial domains. These developments affirm the relevance of models like SFDNN and motivate future extensions that incorporate principled uncertainty quantification and mechanistic constraints.

The remainder of this study is structured as follows. Section~\ref{sec2} introduces the proposed model and its key components. Section~\ref{sec3} outlines the estimation procedure for the spatial autocorrelation term. Section~\ref{sec4} presents the spatial functional deep neural network (SFDNN) approach in detail. In Section~\ref{sec5}, we conduct extensive Monte Carlo experiments to assess the predictive performance of the proposed method. Section~\ref{sec6} provides an in-depth application of the SFDNN model to the Brazilian COVID1-19 dataset, comparing its performance with existing methods. Finally, Section~\ref{sec7} concludes the paper with a discussion of key findings and potential future research directions.

\section{Model, notations, and nomenclature}\label{sec2}

Let $\{ Y_d, \bm{\X}_d(u), \bm{Z}_d: d \in \mathcal{D} \subset \mathbb{R}^q, q \geq 1 \}$ be a stochastic process observed at a set of discrete spatial locations $\mathcal{D} = \{ d_1, \ldots, d_n \}$, where $\mathcal{D}$ may be structured on either a regular or an irregular lattice. The response variable is denoted as $Y_d \in \mathbb{R}$, while the predictor set consists of functional covariates $\bm{\X}_d = \{\X_{1d}, \ldots, \X_{Pd} \}^\top$, with each $\X_{dp}$ belonging to the function space $\mathcal{L}[0,1], \forall p \in \{1, \ldots, P\}$. Additionally, $\bm{Z}_d = (z_{1d}, \ldots, z_{J d})^\top \in \mathbb{R}^{J}$ represents a set of scalar predictors. To simplify notation, we refer to spatial location $d_i$ as $i$. The SSoFRM considered in this study is then formulated as:
\begin{align}
Y &= \beta_0 \bm{1}_n + \rho \bm{W} Y + \int_0^1 \bm{\X}(u) \bm{\beta}(u) \, du + \bm{Z} \bm{\Gamma} + \epsilon, \label{eq:mssofr} \\
&= (\mathbb{I}_n - \rho \bm{W})^{-1} \left\lbrace \beta_0 \bm{1}_n + \int_0^1 \bm{\X}(u) \bm{\beta}(u) \, du + \bm{Z} \bm{\Gamma} + \epsilon \right\rbrace, \nonumber
\end{align}
where $\bm{\beta}(u) = \{\beta_1(u), \ldots, \beta_P(u) \}^\top$ represents the vector of coefficient functions associated with the functional predictors, with each $\beta_p(u) \in \mathcal{L}(0,1)$. The parameter $\bm{\Gamma} = (\gamma_1, \ldots, \gamma_{J})^\top$ corresponds to the regression coefficients for the scalar covariates. The term $\rho \bm{W} Y$ models spatial dependence, where $\bm{W}$ is a known spatial weight matrix and $\rho$ is the associated spatial autocorrelation parameter. The error term $\epsilon$ accounts for random variation and is assumed to follow an appropriate spatially structured covariance model.

The spatial weight matrix $\bm{W} = (w_{ij})_{1 \leq i,j \leq n}$ encodes the strength of spatial relationships between units $i$ and $j$, playing a pivotal role in characterizing the structure of spatial dependence in autoregressive models. Typically, the entries $w_{ij}$ are defined based on a user-specified spatial interaction kernel applied to a pairwise dissimilarity measure, such that $w_{ij} = k_f(d_{ij})$, where  is a monotonically decreasing function, and $d_{ij}$ denotes the spatial distance between units $i$ and $j$, and $k_f(\cdot)$ is a non-negative, non-increasing function satisfying $k_f(0) = 0$ and $k_f(d) \rightarrow 0$ as $d \rightarrow \infty$. Common choices for $k_f$ include inverse distance kernels, Gaussian decay functions, and compactly supported kernels like the bi-square or Epanechnikov kernel. 

In practice, spatial distances $d_{ij}$ can be derived from physical geography (e.g., Euclidean or great-circle distances based on latitude and longitude), network topology (e.g., adjacency or travel time), or application-specific measures that reflect sociopolitical, economic, or environmental connectivity. To promote sparsity and computational tractability, many implementations restrict the neighborhood of each unit to its $k$-nearest neighbors or those within a fixed radius. After applying the kernel function, the resulting weights are often row-normalized to ensure that each row of $\bm{W}$ sums to one, thereby facilitating interpretation and numerical stability in matrix operations. Although symmetry in $\bm{W}$ is frequently assumed for simplicity, the methodology accommodates asymmetric weights when directional interactions or hierarchical influences are intrinsic to the spatial process \citep[see, e.g.,][]{Yu2016, Huang2021}.

As highlighted in Section~\ref{sec1}, a direct application of a link function $g(\cdot)$ to the SSoFRM~\eqref{eq:mssofr} in order to extend the model to a nonlinear setting presents significant challenges. The primary issue stems from the spatial dependence term $\rho \bm{W} Y$: applying a nonlinear transformation to this term before estimating $Y$ disrupts the underlying spatial correlation structure, leading to model misspecification and biased inference. To address this issue, we first estimate the spatial autoregressive parameter $\rho$ using the ML estimator, denoted as $\widehat{\rho}$, ensuring that spatial dependence is appropriately accounted for. Given this estimate, we define the nonlinear spatial scalar-on-function regression model (NSSoFRM) as follows:
\begin{equation}\label{eq:nssofr}
\mathbb{E} \left[ Y \mid \bm{\X}(u), \bm{Z} \right] = g \left\{ (\mathbb{I}_n - \widehat{\rho} \bm{W})^{-1} \left( \beta_0 + \int_0^1 \bm{\X}(u) \bm{\beta}(u) \, du + \bm{Z} \bm{\Gamma} \right) \right\}.
\end{equation}
This formulation ensures that the spatial dependence structure is preserved while enabling a flexible nonlinear transformation via $g(\cdot)$. Consequently, the model retains interpretability and coherence in spatially dependent functional data settings.

\section{Estimation of spatial autocorrelation parameter \texorpdfstring{$\rho$}{rho}}\label{sec3}

We consider the SSoFRM in~\eqref{eq:mssofr}. For each functional predictor $\X_p(u)$, the mean function is defined as $\mu_p(u) = \mathbb{E}[\X_p(u)]$, $u \in [0,1]$. Then, the covariance operator associated with each functional predictor is given by 
\begin{equation*}
C_p(s,u) = \text{Cov} \lbrace \X_p(s), \X_p(u) \rbrace = \mathbb{E}\left[ \lbrace \X_p(s) - \mu_p(s))(\X_p(u) - \mu_p(u) \rbrace \right], \quad s,u \in [0,1].
\end{equation*}
Applying Mercer's Theorem, the covariance function admits the spectral decomposition $C_p(s,u) = \sum_{k=1}^{\infty} \lambda_{pk} \phi_{pk}(s) \phi_{pk}(u)$, where $\lambda_{pk}$ are the eigenvalues in descending order and $\phi_{pk}(u)$ are the corresponding orthonormal eigenfunctions satisfying $\int_{0}^{1} C_p(s,u) \phi_{pk}(u) \, du = \lambda_{pk} \phi_{pk}(s)$. By the Karhunen-Lo\`{e}ve theorem, each function $\X_p(u)$ can be expanded as 
\begin{equation*}
\X_p(u) = \mu_p(u) + \sum_{k=1}^{\infty} \xi_{pk} \phi_{pk}(u), 
\end{equation*}
where the functional principal component (FPC) scores are obtained by projection: $\xi_{pk} = \int_{0}^{1} \{ \X_p(u) - \mu_p(u) \} \phi_{pk}(u) \, du$. To approximate the functional predictor, we retain the first $K_p$ principal components: 
\[
\X_p(u) \approx \mu_p(u) + \sum_{k=1}^{K_p} \xi_{pk} \phi_{pk}(u), 
\]
where $K_p$ is determined according to the explained variance criteria, i.e., the first $K_p$ FPCs for each $\X_p(u)$, $p \in \{1, \ldots, P \}$, capture at least 95\% of the explained variation in the data.

Let $\bm{\xi}_d = (\xi_{11d}, \dots, \xi_{1K_1d}, \xi_{21d}, \dots, \xi_{PK_Pd})^\top \in \mathbb{R}^{K}$ define the concatenated FPC score vector $\bm{\xi}_d$ for location $d$, where $K = \sum_{p=1}^{P} K_p$. This vector captures the functional information in a finite-dimensional form. The scalar covariates $\bm{Z}_d$ are then combined to form the extended design matrix:
\begin{equation*}
\bm{X}_c = \begin{bmatrix} \bm{1}_n & \bm{\Xi} & \bm{Z} \end{bmatrix},
\end{equation*}
where $\bm{\Xi}$ is the matrix of FPC scores and $\bm{Z}$ is the matrix of scalar predictors.

Given the transformed model representation:
\begin{equation*}
Y = \beta_0 \bm{1}_n + \rho \bm{W} Y + \bm{X}_c \bm{\theta} + \epsilon,
\end{equation*}
where $\bm{\theta}$ is the coefficient vector associated with $\bm{X}_c$, and assuming that the error term $\epsilon \sim \mathcal{N}(0, \sigma^2 \mathbb{I}_n)$, the likelihood function is given by
\begin{equation*}
L(\rho, \bm{\theta}, \sigma^2) = \frac{1}{(2\pi \sigma^2)^{n/2}} \vert \mathbb{I}_n - \rho W \vert \exp \left\lbrace -\frac{1}{2\sigma^2} (Y - \rho \bm{W} Y - \bm{X}_c \bm{\theta})^\top (Y - \rho \bm{W} Y - \bm{X}_c \bm{\theta}) \right\rbrace.
\end{equation*}
The corresponding log-likelihood function is:
\begin{equation*}
\ln L(\rho) = -\frac{n}{2} \ln (2\pi \sigma^2) + \ln \vert \mathbb{I}_n - \rho \bm{W} \vert - \frac{1}{2\sigma^2} (Y - \rho \bm{W} Y - \bm{X}_c \bm{\theta})^\top (Y - \rho \bm{W} Y - \bm{X}_c \bm{\theta}).
\end{equation*}

The likelihood is maximized with respect to $\rho$ by solving:
\begin{equation}
\hat{\rho} = \arg\max_{\rho} \left\{ \ln \vert I_n - \rho \bm{W} \vert - \frac{1}{2\sigma^2} (Y - \rho \bm{W} Y - \bm{X}_c \bm{\theta})^\top (Y - \rho \bm{W} Y - \bm{X}_c \bm{\theta}) \right\}.
\end{equation}
This optimization is typically performed using numerical methods such as Newton-Raphson or Broyden–Fletcher–Goldfarb–Shanno (BFGS) algorithm.

\section{Spatial functional deep neural network}\label{sec4}

Neural networks consist of multiple interconnected layers, each comprising a set of computational units known as neurons. These intermediate layers, commonly termed hidden layers, facilitate hierarchical feature extraction by iteratively transforming input data. Specifically, let $n_r$ denote the number of neurons in the $r$-th hidden layer. Each neuron in layer $r$ applies a nonlinear activation function to a weighted sum of the outputs from the preceding layer, thereby enabling the network to capture intricate dependencies and representations within the data. The successive application of these transformations across multiple layers enhances the model’s capacity to approximate highly complex functions.

For a given observation, the output of the first hidden layer, denoted by $v^{(1)}$, is defined as $v^{(1)} = g \left( \omega^{(1)} x + b^{(1)} \right)$, where $x$ is a vector of $J$ covariates, $\omega^{(1)}$ represents an $n_1 \times J$ weight matrix, and $b^{(1)}$ is the corresponding bias term for the first hidden layer. The function $g(\cdot)$ serves as the activation function, introducing a nonlinear transformation to the affine combination of the input features \citep{Tibshirani2009, Thind2023}. The resulting transformed vector $v^{(1)}$, of dimension $n_1$, is subsequently propagated to the next layer, enabling the network to extract hierarchical representations from the data.

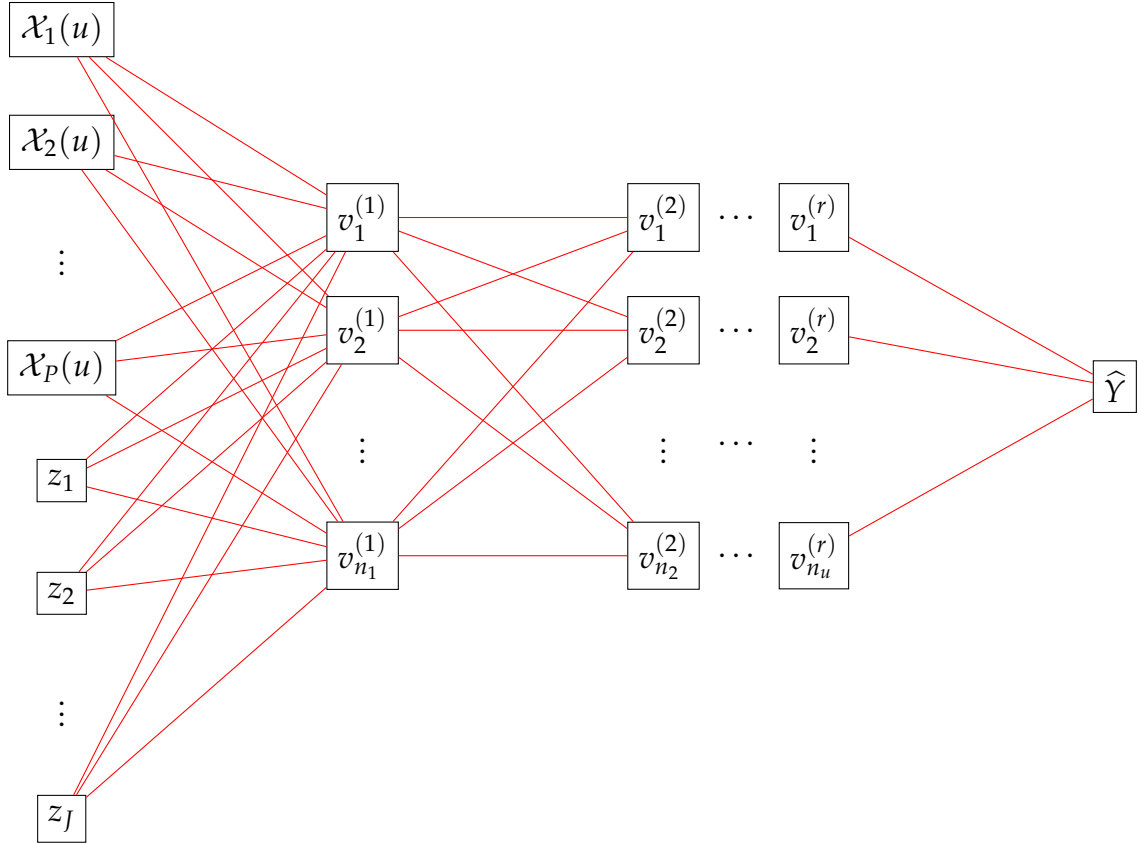
\begin{figure}[!htb]
\centering
\begin{tikzpicture}[every node/.style={align=center}, node distance=1.5cm]

\node[draw, rectangle] (x1) at (0, 3) {$\X_1(u)$};
\node[draw, rectangle] (x2) at (0, 1.5) {$\X_2(u)$};
\node[draw=none] (dots1) at (0, 0) {$\vdots$};
\node[draw, rectangle] (xk) at (0, -1.5) {$\X_P(u)$};

\node[draw, rectangle] (z1) at (0, -3.0) {$z_1$};
\node[draw, rectangle] (z2) at (0, -4.5) {$z_2$};
\node[draw=none] (dots2) at (0, -6.0) {$\vdots$};
\node[draw, rectangle] (zj) at (0, -7.5) {$z_{J}$};

\node[draw, rectangle] (h11) at (4, 0.5) {$v_{1}^{(1)}$};
\node[draw, rectangle] (h12) at (4, -1.0) {$v_{2}^{(1)}$};
\node[draw=none] (dots_h1) at (4, -2.5) {$\vdots$};
\node[draw, rectangle] (h1u) at (4, -4.0) {$v_{n_1}^{(1)}$};

\node[draw, rectangle] (h21) at (8, 0.5) {$v_{1}^{(2)}$};
\node[draw, rectangle] (h22) at (8, -1.0) {$v_{2}^{(2)}$};
\node[draw=none] (dots_h2) at (8, -2.5) {$\vdots$};
\node[draw, rectangle] (h2u) at (8, -4.0) {$v_{n_2}^{(2)}$};

\node[draw=none] (dots_row1) at (9, 0.5) {$\dots$};
\node[draw=none] (dots_row2) at (9, -1.0) {$\dots$};
\node[draw=none] (dots_row3) at (9, -2.5) {$\dots$};
\node[draw=none] (dots_row4) at (9, -4.0) {$\dots$};

\node[draw, rectangle] (hu1) at (10, 0.5) {$v_{1}^{(r)}$};
\node[draw, rectangle] (hu2) at (10, -1.0) {$v_{2}^{(r)}$};
\node[draw=none] (dots_hu) at (10, -2.5) {$\vdots$};
\node[draw, rectangle] (huu) at (10, -4.0) {$v_{n_u}^{(r)}$};

\node[draw, rectangle] (yhat) at (14, -1.75) {$\widehat{Y}$};

\foreach \i in {x1,x2,xk,z1,z2,zj} {
    \foreach \j in {h11,h12,h1u} {
        \draw[red] (\i) -- (\j);
    }
}

\foreach \i in {h11,h12,h1u} {
    \foreach \j in {h21,h22,h2u} {
        \draw[red] (\i) -- (\j);
    }
}

\foreach \i in {hu1,hu2,huu} {
    \draw[red] (\i) -- (yhat);
}
\end{tikzpicture}
\caption{Schematic architecture of the proposed Spatial Functional Deep Neural Network (SFDNN), which integrates both functional and scalar predictors to model a scalar response with spatial dependencies. The functional covariates $\X_p(u)$, are first projected onto a finite-dimensional space using basis function expansions (e.g., B-splines), producing scalar summaries that are fed into the network. Scalar covariates $z_j$, for $j \in \{1, \ldots, J \}$, are directly incorporated without transformation. These inputs enter the first hidden layer, where each neuron computes a nonlinear transformation of a weighted combination of functional and scalar inputs adjusted for spatial dependence using the matrix inverse $(\mathbb{I}_n - \widehat{\rho} \bm{W})^{-1}$. The activations then propagate through multiple hidden layers, capturing hierarchical nonlinear interactions. The network outputs a single scalar prediction, $\widehat{Y}$, which corresponds to the expected value of the scalar response variable.
}
\label{fig:fig_1}
\end{figure}

In the NSSoFRM considered in this study, the covariates include infinite-dimensional functional predictors $\X_p(u)$, for $p \in \{1, \ldots, P \}$. Unlike classical neural networks, where weights are finite-dimensional matrices, FDNN must assign weights continuously over the entire domain of each functional covariate. Consequently, the weight functions themselves are infinite-dimensional to ensure that the contribution of the functional predictors is properly accounted for at every point along their domains. To address this, we adopt the FDNN approach proposed by \cite{Thind2023}, as illustrated in Figure~\ref{fig:fig_1}.

For the case of a single functional covariate, let $\omega_\ell(u)$, for $\ell \in \{1, \ldots, n_1 \}$, represent the functional weight associated with the $\ell$-th neuron in the first hidden layer (see Figure~\ref{fig:fig_1}). The corresponding neuron activation is defined as:
\begin{equation}\label{eq:fneur}
v_\ell^{(1)} = g \left[ \left(\mathbb{I}_n - \widehat{\rho} \bm{W} \right)^{-1} \left\lbrace \int_0^1 \omega_\ell(u) \X(u) \, du \right\rbrace + b_\ell^{(1)} \right],
\end{equation}
where $g(\cdot)$ denotes the activation function, $\bm{W}$ is the spatial weight matrix, and $b_\ell^{(1)}$ is the bias term for the first hidden layer. Note that the superscript on the functional weight $\omega(u)$ is omitted, as it pertains specifically to the first layer of the network.

The functional weights in the model are represented as linear combinations of basis functions, allowing for an efficient finite-dimensional parameterization. Specifically, we express each functional weight as:
\begin{equation}\label{eq:bspl}
\omega_{\ell}(u) = \sum_{m=1}^{M} c_{\ell m} \psi_{\ell m}(u) = \bm{c}_\ell^\top \bm{\psi}_\ell(u),
\end{equation}
where $\bm{\psi}_\ell(u) = \{ \psi_{\ell 1}(u), \ldots, \psi_{\ell M}(u) \}^\top$ is the vector of basis functions, and $\bm{c}_\ell = (c_{\ell 1}, \ldots, c_{\ell M})^\top$ denotes the corresponding vector of basis coefficients. The parameter $M$ represents the number of basis functions and serves as a hyperparameter in the FDNN model. In this study, we employ the B-spline basis expansion \citep{deboor2001}, given its flexibility and computational efficiency. The coefficients $c_{\ell m}$ are initialized within the network and subsequently optimized during training.

Substituting the basis representation from~\eqref{eq:bspl} into the neuron activation function in~\eqref{eq:fneur}, we obtain:
\begin{align*}
v_\ell^{(1)} &= g \left[ \left(\mathbb{I}_n - \widehat{\rho} \bm{W} \right)^{-1} \left\lbrace \int_0^1 \omega_\ell(u) \X(u) \, du \right\rbrace + b_\ell^{(1)} \right], \\
&= g \left[ \left(\mathbb{I}_n - \widehat{\rho} \bm{W} \right)^{-1} \left\lbrace \int_0^1 \sum_{m=1}^{M} c_{\ell m} \psi_{\ell m}(u) \X(u) \, du \right\rbrace + b_\ell^{(1)} \right], \\
&= g \left[ \left(\mathbb{I}_n - \widehat{\rho} \bm{W} \right)^{-1} \left\lbrace \sum_{m=1}^{M} c_{\ell m} \int_0^1 \psi_{\ell m}(u) \X(u) \, du \right\rbrace + b_\ell^{(1)} \right],
\end{align*}
where the integral term $\int_0^1 \psi_{\ell m}(u) \X(u) \, du$ is numerically approximated using appropriate quadrature methods. Notably, the evaluation of each neuron in~\eqref{eq:fneur} results in a scalar output, enabling subsequent layers to follow standard deep learning architectures, such as feedforward networks.

Extending the previous formulation to the NSSoFRM, where the model incorporates both $P$ functional covariates and $J$ scalar covariates, the activation of the $\ell$-th neuron in the first hidden layer is given by:
\begin{align*}
v_\ell^{(1)} &= g \left[ \left(\mathbb{I}_n - \widehat{\rho} \bm{W} \right)^{-1} \left\lbrace \sum_{p=1}^P \int_0^1 \omega_{\ell p}(u) \X_p(u) \, du + \sum_{j=1}^{J} \omega_{\ell j}^{(1)} z_j \right\rbrace + b_\ell^{(1)} \right], \\
&= g \left[ \left(\mathbb{I}_n - \widehat{\rho} \bm{W} \right)^{-1} \left\lbrace \sum_{p=1}^P \int_0^1 \sum_{m=1}^{M_P} c_{\ell p m} \psi_{\ell p m}(u) \X_p(u) \, du + \sum_{j=1}^{J} \omega_{\ell j}^{(1)} z_j \right\rbrace + b_\ell^{(1)} \right], \\
&= g \left[ \left(\mathbb{I}_n - \widehat{\rho} \bm{W} \right)^{-1} \left\lbrace \sum_{p=1}^P \sum_{m=1}^{M_P} c_{\ell p m} \int_0^1 \psi_{\ell p m}(u) \X_p(u) \, du + \sum_{j=1}^{J} \omega_{\ell j}^{(1)} z_j \right\rbrace + b_\ell^{(1)} \right],
\end{align*}
where $\omega_{\ell j}^{(1)}$ denotes the weight associated with the $j$-th scalar covariate $z_j$. The weight function corresponding to the $p$-th functional covariate, $\omega_{\ell p}(u)$, is parameterized using a B-spline basis expansion $\omega_{\ell p}(u) = \sum_{m=1}^{M_P} c_{\ell p m} \psi_{\ell p m}(u)$, where $\psi_{\ell p m}(u)$ represents the basis expansion functions, and $c_{\ell p m}$ denotes the corresponding expansion coefficients for $p \in \{1, \ldots, P \}$ and $m \in \{1, \ldots, M_P \}$. This formulation enables the functional weights to be learned efficiently within the neural network framework, facilitating the extraction of meaningful representations from both functional and scalar inputs.

After propagating the input through the initial neurons in the first hidden layer and computing activations across subsequent layers, the network ultimately produces a single-dimensional output. To evaluate the performance of the model, we employ a loss function that quantifies the discrepancy between predicted and observed values. A common choice is the mean squared error (MSE), defined as $\varphi(\Theta) = \sum_{i=1}^N \{Y_i - \widehat{Y}_i(\Theta) \}^2$, where $\Theta$ represents the set of parameters characterizing the neural network, and $\widehat{Y}_i(\Theta)$ denotes the predicted output from the functional deep neural network (FDNN) for the $i$-th observation. This loss function serves as an objective criterion for optimizing the model parameters during training, enabling the network to learn meaningful representations from both functional and scalar inputs.

\subsection{FDNN training and parameter tuning}

The training procedure for the FDNN follows a gradient-based optimization framework in which all functional and scalar weight parameters, as well as neuron biases, are learned via backpropagation \citep{Rumelhart1985}. The network parameters are adjusted using the Adam optimization algorithm \citep{Kingma2014}. Specifically, once the spatial autocorrelation parameter $\rho$ is estimated via maximum likelihood, it is treated as fixed. The matrix inversion $(\mathbb{I}_n - \widehat{\rho} \bm{W})^{-1}$  is then used to spatially adjust the covariates before feeding them into the network. Importantly, only the parameters of the FDNN—namely, the functional basis coefficients $\{c_{\ell p m} \}$, scalar weights $\{\omega_{\ell j}^{(1)} \}$, and layer-specific biases $\{b_\ell^{(r)} \}$—are optimized during neural network training. The spatial parameter $\rho$ is not updated during the FDNN phase, thereby ensuring stability of the spatial structure.

Let $\bigtriangleup \Theta$ denote the gradient set for all learnable parameters, expressed as:
\begin{equation*}
\bigtriangleup \Theta = \left\{ \bigcup_{p=1}^{P} \bigcup_{m=1}^{M_P} \bigcup_{\ell=1}^{n_1} \frac{\partial \varphi}{\partial c_{\ell p m}}, \quad
\bigcup_{r=1}^{R} \bigcup_{j=1}^{J_r} \bigcup_{\ell=1}^{n_r} \frac{\partial \varphi}{\partial w_{\ell j r}}, \quad
\bigcup_{r=1}^{R} \bigcup_{\ell=1}^{n_r} \frac{\partial \varphi}{\partial b_{\ell r}} \right\}.
\end{equation*}
Since these gradients are computed for each individual training observation, the optimization procedure seeks to adjust parameters across the entire dataset. Instead of updating the parameters based on the full training set at each step— which would be computationally expensive— we employ a stochastic approach by selecting a mini-batch of $N_b$ randomly sampled observations. The estimated gradient over this mini-batch is given by $\widetilde{a} = \frac{1}{N_b} \sum_{i=1}^{N_b} \frac{\partial \varphi}{\partial a_i}$. The corresponding parameter update is performed as $a \gets a - \zeta \widetilde{a}$, where $\zeta$ represents the learning rate, a hyperparameter controlling the step size in the optimization process \citep{Ruder2016}. This procedure is repeated iteratively over successive mini-batches until all subsets of the dataset are processed, completing one full training cycle, or epoch. The number of epochs is treated as a hyperparameter and tuned based on model performance and convergence criteria. 

The tuning process for the functional deep neural network (FDNN) is outlined in Algorithm~\ref{alg1}. Given the presence of multiple hyperparameters in the network, an appropriate selection strategy is necessary to identify optimal values. A common approach involves systematically exploring a predefined set of candidate values for each hyperparameter and evaluating their performance using cross-validation \citep{Tibshirani2009, Thind2023}.
\begin{algorithm}
\caption{\textbf{Training procedure for functional neural networks}}\label{alg1}
\begin{algorithmic}
\Require Functional inputs $\X_p(u)$ for $p \in \{1, \ldots, P \}$ and scalar inputs $z_j$ for $j \in \{1, \ldots, J \}$.  
Set hyperparameters: learning rate $\zeta$, number of hidden layers, neurons per layer, activation functions, weight decay, validation split, basis expansion size for functional weights, domain of functional weights, number of training epochs, mini-batch size, early stopping criteria, and threshold for convergence.
\Ensure Optimized neural network parameters $\Theta$.
\State \textbf{Step 1: Initialize Hyperparameters}
\State Define the stopping criterion $\tau > 0$, representing the minimum loss improvement threshold.
\State Set iteration counter $\eta = 0$ and initialize loss difference $\delta = 0$.
\State \textbf{Step 2: Training Process}
\While{ $\delta < \tau$}
    \State \textbf{(a) Forward Propagation:}
    \State \quad Pass functional and scalar covariates $\X_p(u)$ and $z_j$ into the first hidden layer.
    \State \quad Approximate integral terms for functional inputs using basis expansion:
    \begin{equation*}
    \int_0^1 \psi_{\ell m}(u) \X_p(u) \, du \approx \widetilde{\psi}_{\ell m}
    \end{equation*}
    \State \quad Compute neuron activations:
    \begin{equation*}
    E = g \left\lbrace (\mathbb{I}_n - \widehat{\rho} \bm{W} )^{-1} \left(\sum_{p=1}^{P} \sum_{m=1}^{M_P} c_{\ell p m} \widetilde{\psi}_{\ell p m} + \sum_{j=1}^{J} \omega_{\ell j}^{(1)} z_j \right) + b_{\ell}^{(1)} \right\rbrace
    \end{equation*}
    \State \quad Pass activations through the remaining layers of the network.
    \State \quad Compute loss function $\varphi(\Theta)$.
    \State \textbf{(b) Backpropagation:}
    \State \quad Compute gradients $\bigtriangleup \Theta$ for all network parameters.
    \For{each parameter $a \in \Theta$}
        \State Update parameter using gradient descent:
        \begin{equation*}
        a \gets a - \zeta \widetilde{a}
        \end{equation*}
    \EndFor
    \State \textbf{(c) Evaluate Stopping Criteria:}
    \If{ $\eta = 0$}
        \State Initialize loss improvement measure: $\delta = \varphi(\Theta)$.
    \Else
        \State Update loss improvement: $\delta = \vert \delta - \varphi(\Theta) \vert$.
    \EndIf
    \State Increment iteration counter: $\eta \gets \eta + 1$.
\EndWhile
\State \textbf{Step 3: Return Optimized Parameters}
\State Return the final set of trained parameters $\Theta$.
\end{algorithmic}
\end{algorithm}

To implement this, we construct a search grid encompassing all possible hyperparameter combinations. For each candidate configuration, we perform $K$-fold cross-validation, partitioning the dataset into $K$ non-overlapping subsets. At each iteration, the model is trained on $K-1$ subsets while the remaining subset serves as the validation set. The predictive performance is assessed using the mean squared prediction error (MSPE), computed as $\text{MSPE} = \frac{1}{n} \sum_{k=1}^{K} \sum_{i \in \mathcal{S}_b} (\widehat{Y}_i^{(-b)} - Y_i)^2$, where $\mathcal{S}_b$ denotes the $b$-th validation subset, and $\widehat{Y}_i^{(-b)}$ represents the predicted response for observation $i$ obtained by training the FDNN on the remaining $K-1$ folds. The size of each validation subset \( \mathcal{S}_b \) depends on the total number of data points and the number of folds chosen. The optimal hyperparameter configuration is determined as the one yielding the lowest cross-validated MSPE. 

\section{Simulation studies}\label{sec5}

To assess the predictive accuracy of the proposed method (``SFDNN'', hereafter), we conduct a series of Monte Carlo simulations. The performance of SFDNN is benchmarked against two alternative approaches: the classical FDNN and the ML-based model. In the ML-based approach, functional predictors are projected onto a lower-dimensional space using FPC analysis, followed by parameter estimation via ML. An example \Rlogo \ script for the simulation studies can be accessed at the following GitHub repository: \url{https://github.com/mervebasaran/SpatialFunctionDNN}.

The data generation process follows an extended version of the framework introduced by \cite{BSM2025}. We simulate three functional covariates, denoted as $\bm{X}(u)$ = $\{\X_1(u)$, $\X_2(u)$, $\X_3(u)\}^{\top}$, alongside three scalar predictors, $\bm{Z} = (z_1, z_2, z_3)^\top$. Each functional predictor, $\X_p(u)$, for $p \in \{1,2,3\}$, is observed at 101 equally spaced points over the domain $[0,1]$ and is generated according to:
\begin{equation*}
\X_p(u) = \sum_{j=1}^{5} \kappa_j \upsilon_j(u),
\end{equation*}
where $\kappa_j \sim \mathcal{N}(0, 4 j^{-3/2})$ and the basis functions are defined as $\upsilon_j(u) = \sin(j \pi u) - \cos(j \pi u)$. The regression coefficient functions corresponding to the functional predictors are given by:
\begin{equation*}
\beta_1(u) = \sin(2 \pi u), \quad \beta_2(u) = \cos(2 \pi u), \quad \beta_3(u) = 2 \sin(2 \pi u).
\end{equation*}
The scalar covariates are sampled from a standard normal distribution, $z_j \sim \mathcal{N}(0,1)$ for $j \in \{1,2,3\}$, with corresponding regression coefficients set as $\bm{\Gamma} = (1.25, -2, 2.15)^\top$.

To incorporate spatial dependence, we construct a spatial weight matrix $\bm{W}$ based on inverse distance weighting. The elements of $\bm{W}$ are defined as:
\begin{equation*}
w_{i i^{\prime}} = \frac{1}{1 + | i - i^{\prime} |}, \quad i \neq i^{\prime},
\end{equation*}
with diagonal elements set to zero, i.e., $w_{ii} = 0$ for all $i$. To ensure proper scaling, the matrix is row-normalized as:
\begin{equation*}
\bm{W}_{i.} = \frac{w_{i i^{\prime}}}{\sum_{i^{\prime}=1}^{n} w_{i i^{\prime}}}, \quad \forall i.
\end{equation*}

The response variable in our simulation study is generated according to the following SSoFRM:
\begin{equation*}
Y = (\mathbb{I}_n - \rho \bm{W})^{-1} \left\lbrace \beta_0 \bm{1}_n + \int_0^1 \bm{X}(u) \bm{\beta}(u) \, du + \bm{Z} \bm{\Gamma} + (\mathbb{I}_n - \rho \bm{W})^{-1} \epsilon \right\rbrace .
\end{equation*}
We consider three scenarios for the error distribution; 1) $\epsilon \sim \mathcal{N}(0,1)$, 2) $\epsilon \sim t_3$ ($t$-distribution with non-centrality parameter, and 3) $\epsilon \sim \exp(1)$. The heavy-tailed $t_3$-distributed errors introduce non-Gaussian noise with occasional extreme values, leading to deviations from the standard linear assumption and increasing the complexity of model estimation. Similarly, the exponentially distributed errors $\exp(1)$ are positively skewed, inducing asymmetry in the response variable and creating nonlinear patterns that challenge traditional modeling approaches.

\begin{figure}[!htb]
\centering
\includegraphics[width=5.8cm]{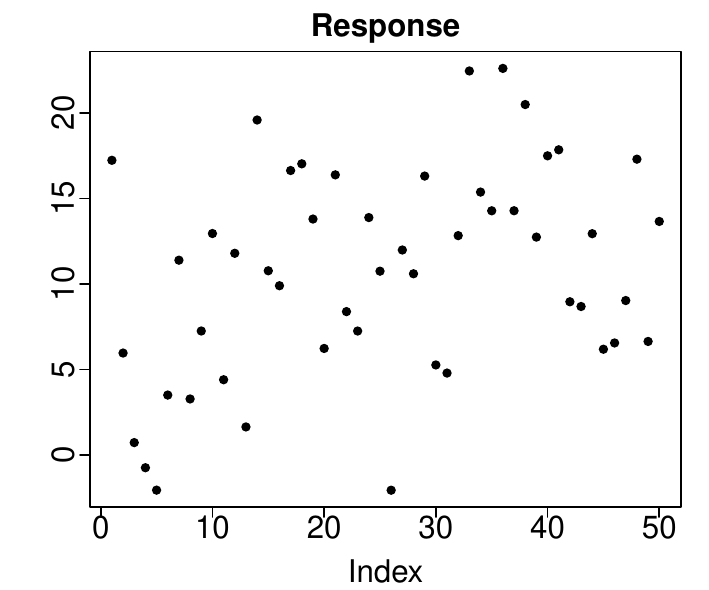}
\quad
\includegraphics[width=5.8cm]{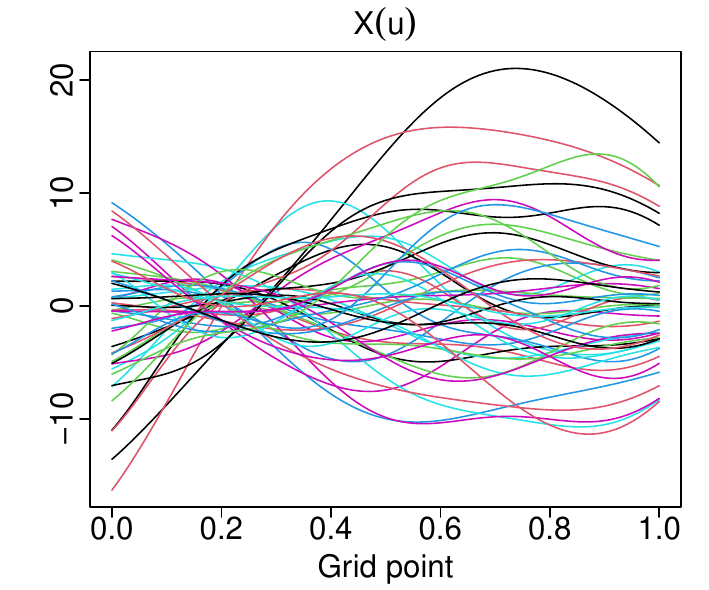}
\quad
\includegraphics[width=5.8cm]{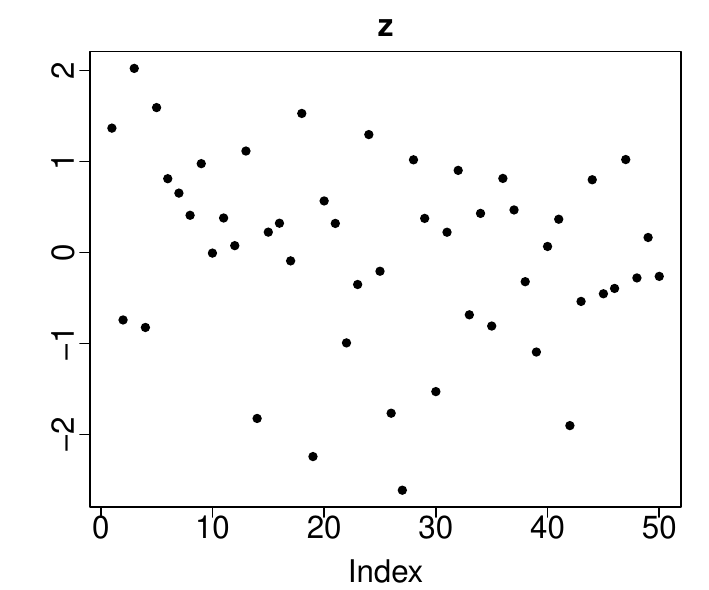}
\\  
\includegraphics[width=5.8cm]{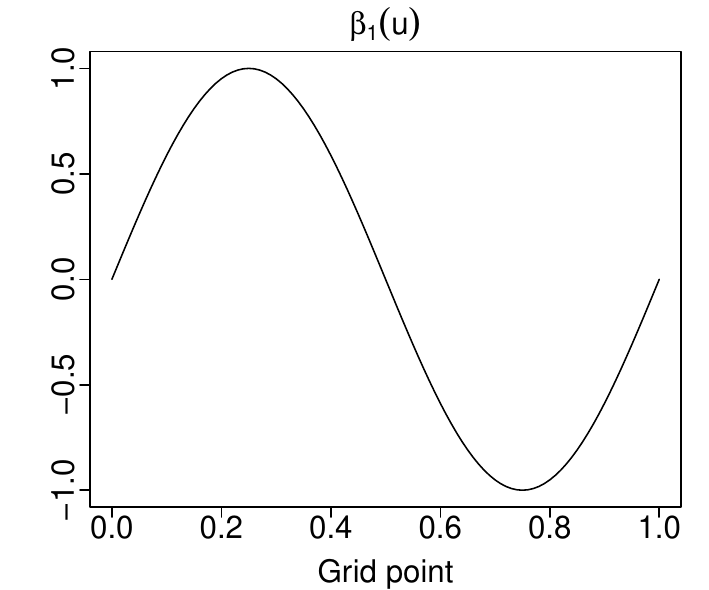}
\quad
\includegraphics[width=5.8cm]{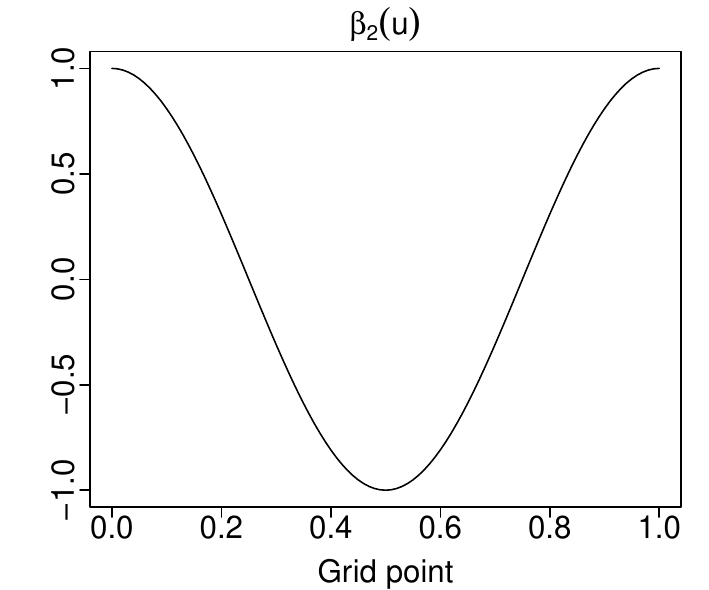}
\quad
\includegraphics[width=5.8cm]{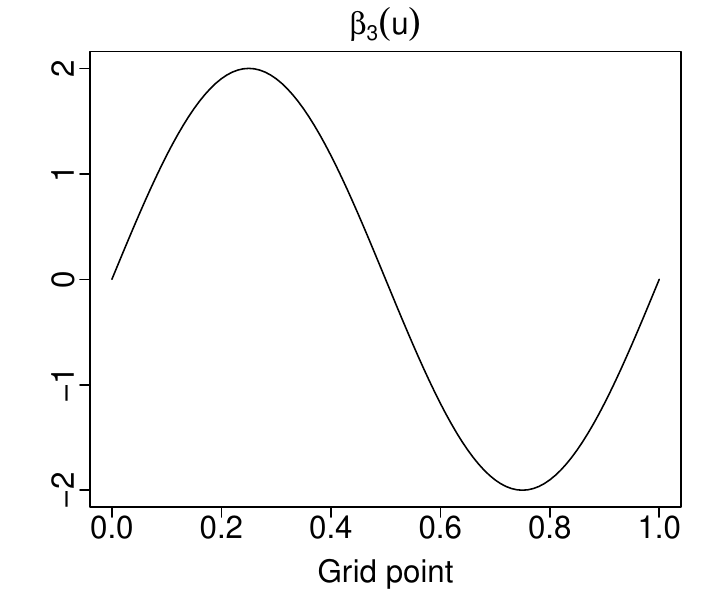}
\caption{\small{Graphs illustrating 50 generated sample observations for the scalar response (top-left panel), functional covariates (top-middle panel; as all three functional covariates exhibit similar structures, only one is shown), and scalar covariates (top-right panel; similarly, only one of the three scalar covariates is presented). The bottom panels display the true regression coefficient functions. The data are generated when $\epsilon \sim \exp(1)$ with $\rho = 0.9$}.}\label{fig:Fig_2}
\end{figure}

To assess the impact of spatial dependence, we examine three distinct values for the spatial autocorrelation parameter: $\rho \in \{0.1, 0.5, 0.9\}$, representing weak, moderate, and strong spatial effects, respectively. Across different experimental settings, both the FDNN and its spatial variant SFDNN are expected to outperform the traditional ML-based approach, particularly when nonlinear transformations are present (i.e., under the exponential and sigmoidal link functions). Moreover, the proposed SFDNN is anticipated to provide superior predictive accuracy in scenarios where spatial dependencies are more pronounced, specifically when $\rho$ takes moderate to high values ($\rho = 0.5, 0.9$). Figure~\ref{fig:Fig_2} provides a visual representation of the simulated dataset along with the corresponding regression coefficient functions under the sigmoidal link transformation, specifically for the case where the spatial dependence parameter is set to $\rho = 0.9$.

In the simulation experiments, we evaluate model performance under different training sample sizes, specifically considering $n_{\text{train}} \in \{100, 250, 500\}$. For each scenario, we generate an independent test set of size $n_{\text{test}} = 1000$ to assess generalization performance. The models are first trained on the generated datasets, and their predictive accuracy is subsequently evaluated using the test samples. 

To quantitatively assess predictive performance, we employ two key evaluation metrics: the mean squared error (MSE) and the coefficient of determination $R^2$ for both training and test sets. Specifically, in-sample prediction accuracy is measured using the MSE and $R^2$ on the training data, while out-of-sample performance is assessed using the MSPE and the test sample coefficient of determination ( $R^2_{\text{test}}$). These metrics are computed as follows:
\begin{align*}
\text{MSE} &= \frac{1}{n_{\text{train}}} \sum_{i=1}^{n_{\text{train}}} (Y_i - \widehat{Y}_i)^2, \qquad R^2 = 1 - \frac{\sum_{i=1}^{n_{\text{train}}} (Y_i - \widehat{Y}_i)^2}{\sum_{i=1}^{n_{\text{train}}} (Y_i - \overline{Y})^2}, \\
\text{MSPE} &= \frac{1}{n_{\text{test}}} \sum_{i=1}^{n_{\text{test}}} (Y_i^* - \widehat{Y}_i^*)^2, \qquad R^2_{\text{test}} = 1 - \frac{\sum_{i=1}^{n_{\text{test}}} (Y_i^* - \widehat{Y}_i^*)^2}{\sum_{i=1}^{n_{\text{test}}} (Y_i^* - \overline{Y}^*)^2}.
\end{align*}
Here, the superscript $^*$ indicates that the observations belong to the test dataset, while $\overline{Y}$ and $\overline{Y}^*$ represent the mean response values in the training and test sets, respectively. To ensure a valid assessment of generalization performance, the training and test samples are constructed so as not to contain overlapping observations.

The simulation results are presented in Tables~\ref{tab:tab_1}-\ref{tab:tab_3}. Under Gaussian errors, i.e., Table~\ref{tab:tab_1}, all three models demonstrate strong predictive performance in low spatial dependence scenarios ($\rho = 0.1$), with relatively small differences in MSPE and $R^2_{\text{test}}$. However, as spatial autocorrelation increases ($\rho = 0.5$ and $\rho = 0.9$), the ML-based model exhibits a substantial rise in MSPE, reflecting its inability to adequately capture spatial dependencies. The FDNN, while more robust than ML, also shows a notable degradation in predictive accuracy under high spatial dependence. In contrast, the SFDNN consistently maintains the lowest MSPE across all spatial correlation levels, achieving the highest $R^2_{\text{test}}$, which underscores the effectiveness of incorporating explicit spatial dependencies within a deep learning framework. Notably, when $\rho = 0.9$, the MSPE of ML is nearly three times that of SFDNN, confirming the necessity of spatially-aware modeling for complex functional data structures.
\begin{small}
\begin{center}
\tabcolsep 0.045in
\begin{longtable}{@{}lccccccccccccc@{}} 
\caption{\small{Comparison of ML, FDNN, and SFDNN based on MSE, MPSE, and $R^2$ metrics computed from 250 Monte Carlo simulations where the error distribution follows $\epsilon \sim \mathcal{N}(0,1)$. Standard errors are in parentheses. Performance metrics are computed under three different training sample sizes ($n_{\text{train}}$) and three different spatial autocorrelation parameter ($\rho$)}.}\label{tab:tab_1} \\
\toprule
 &  & \multicolumn{4}{c}{ML} & \multicolumn{4}{c}{FDNN} & \multicolumn{4}{c}{SFDNN} \\
\midrule
 $n_{\text{train}}$  & $\rho$ & MSE & $R^2$ & MPSE & $R^2_{\text{test}}$ & MSE & $R^2$ & MPSE & $R^2_{\text{test}}$ & MSE & $R^2$ & MPSE & $R^2_{\text{test}}$ \\
\midrule
\endfirsthead
\toprule
&  & \multicolumn{4}{c}{ML} & \multicolumn{4}{c}{FDNN} & \multicolumn{4}{c}{SFDNN} \\
 $n_{\text{train}}$ &  $\rho$ & MSE & $R^2$ & MPSE & $R^2_{\text{test}}$ & MSE & $R^2$ & MPSE & $R^2_{\text{test}}$ & MSE & $R^2$ & MPSE & $R^2_{\text{test}}$ \\
\midrule
\endhead
\endfoot
\endlastfoot
 100 & 0.1 & 0.988 & 0.967 & 2.344 & 0.936 & 0.966 & 0.968 & 1.438 & 0.953 & 0.963 & 0.968 & 1.516 & 0.952 \\ 
 & & (0.200) & (0.007) & (1.181) & (0.029) & (0.177) & (0.006) & (0.134) & (0.005) & (0.164) & (0.005) & (0.209) & (0.006) \\
 \cmidrule{2-14}
  & 0.5 & 1.045 & 0.961 & 3.454 & 0.916 & 1.142 & 0.960 & 2.124 & 0.946 & 0.985 & 0.964 & 1.629 & 0.950 \\ 
 & & (0.230) & (0.010) & (3.672) & (0.102) & (0.224) & (0.007) & (0.628) & (0.007) & (0.170) & (0.007) & (0.322) & (0.008) \\
 \cmidrule{2-14}
  & 0.9 & 1.193 & 0.964 & 37.165 & 0.928 & 5.672 & 0.870 & 44.441 & 0.857 & 3.656 & 0.919 & 13.782 & 0.884 \\ 
 & & (0.290) & (0.010) & (30.278) & (0.046) & (5.880) & (0.100) & (44.583) & (0.099) & (5.796) & (0.102) & (13.919) & (0.108) \\
\midrule
 250 & 0.1 & 1.098 & 0.962 & 2.625 & 0.918 & 1.081 & 0.963 & 1.300 & 0.955 & 1.069 & 0.964 & 1.315 & 0.955 \\ 
 & & (0.109) & (0.005) & (2.043) & (0.067) & (0.097) & (0.006) & (0.116) & (0.005) & (0.087) & (0.006) & (0.099) & (0.004) \\
 \cmidrule{2-14}
  & 0.5 & 1.150 & 0.960 & 2.505 & 0.935 & 1.278 & 0.957 & 1.669 & 0.953 & 1.084 & 0.963  & 1.384 & 0.957 \\ 
 & & (0.211) & (0.008) & (1.694) & (0.056) & (0.177) & (0.007) & (0.348) & (0.006) & (0.151) & (0.007) & (0.180) & (0.006) \\
 \cmidrule{2-14}
  & 0.9 & 1.291 & 0.959 & 16.232 & 0.929 & 2.388 & 0.930 & 29.068 & 0.922 & 1.225 & 0.962 & 8.219 & 0.948 \\ 
 & & (0.145) & (0.005) & (16.227) & (0.039) & (1.159) & (0.027) & (38.912) & (0.013) & (0.112) & (0.006) & (12.748) & (0.015) \\
\midrule
500 & 0.1 & 1.123 & 0.962 & 3.096 & 0.904 & 1.001 & 0.966 & 1.215 & 0.960 & 1.020 & 0.966 & 1.244 & 0.960 \\ 
& & (0.172) & (0.006) & (4.142) & (0.119) & (0.084) & (0.003) & (0.117) & (0.005) & (0.118) & (0.004) & (0.105) & (0.005) \\
\cmidrule{2-14}
  & 0.5 & 1.214 & 0.959 & 1.875 & 0.947 & 1.173 & 0.961 & 1.497 & 0.956 & 1.040 & 0.966 & 1.185 & 0.962 \\ 
 & & (0.166) & (0.005) & (0.840) & (0.018) & (0.119) & (0.003) & (0.328) & (0.006) & (0.098) & (0.003) & (0.136) & (0.004) \\
 \cmidrule{2-14}
  & 0.9 & 1.244 & 0.960 & 15.892 & 0.949 & 2.023 & 0.937 & 23.730 & 0.919 & 1.109 & 0.965 & 4.108 & 0.957 \\ 
 & & (0.160) & (0.004) & (15.210) & (0.029) & (0.495) & (0.011) & (22.085) & (0.026) & (0.106) & (0.005) & (4.792) & (0.009) \\
\bottomrule
\end{longtable}
\end{center}
\end{small}

\vspace{-.3in}

Under the heavy-tailed error scenario, i.e., Table~\ref{tab:tab_2}, the ML and FDNN approaches experience severe performance deterioration, particularly at higher spatial dependence levels, as evidenced by the sharp increase in MSPE. The impact of extreme values is especially pronounced for ML, which lacks the flexibility to adapt to heavy-tailed distributions. The FDNN model provides a moderate improvement but remains susceptible to error inflation due to its inability to explicitly model spatial dependencies. The SFDNN, however, demonstrates remarkable resilience against heavy-tailed noise, achieving significantly lower MSPE and higher $R^2_{\text{test}}$, particularly when $\rho = 0.9$. 

\begin{small}
\begin{center}
\tabcolsep 0.039in
\begin{longtable}{@{}lccccccccccccc@{}} 
\caption{\small{Comparison of ML, FDNN, and SFDNN based on MSE, MPSE, and $R^2$ metrics computed from 250 Monte Carlo simulations where the error distribution follows $\epsilon \sim t_3$ ($t$-distribution with non-centrality parameter. Standard errors are in parentheses. Performance metrics are computed under three different training sample sizes ($n_{\text{train}}$) and three different spatial autocorrelation parameter~($\rho$)}.}\label{tab:tab_2} \\
\toprule
 &  & \multicolumn{4}{c}{ML} & \multicolumn{4}{c}{FDNN} & \multicolumn{4}{c}{SFDNN} \\
\midrule
$n_{\text{train}}$  & $\rho$ & MSE & $R^2$ & MPSE & $R^2_{\text{test}}$ & MSE & $R^2$ & MPSE & $R^2_{\text{test}}$ & MSE & $R^2$ & MPSE & $R^2_{\text{test}}$ \\
\midrule
\endfirsthead
\toprule
  &  & \multicolumn{4}{c}{ML} & \multicolumn{4}{c}{FDNN} & \multicolumn{4}{c}{SFDNN} \\
$n_{\text{train}}$ &  $\rho$ & MSE & $R^2$ & MPSE & $R^2_{\text{test}}$ & MSE & $R^2$ & MPSE & $R^2_{\text{test}}$ & MSE & $R^2$ & MPSE & $R^2_{\text{test}}$ \\
\midrule
\endhead
\endfoot
\endlastfoot
100 & 0.1 & 3.000 & 0.905 & 4.158 & 0.879 & 3.018 & 0.907 & 4.028 & 0.877 & 3.047 & 0.906 & 4.374 & 0.870 \\ 
& & (1.549) & (0.037) & (1.265) & (0.035) & (1.724) & (0.041) & (1.210) & (0.036) & (1.687) & (0.039) & (1.373) & (0.041) \\
\cmidrule{2-14}
  & 0.5 & 2.885 & 0.900 & 4.957 & 0.883 & 3.942 & 0.873 & 6.054 & 0.860 & 2.920 & 0.900 & 4.247 & 0.884 \\ 
 & & (2.120) & (0.058) & (1.558) & (0.029) & (4.459) & (0.121) & (5.025) & (0.096) & (2.132) & (0.058) & (0.955) & (0.023) \\
 \cmidrule{2-14}
  & 0.9 & 3.043 & 0.914 & 38.242 & 0.818 & 6.501 & 0.857 & 45.208 & 0.822 & 3.053 & 0.915 & 22.641 & 0.873 \\ 
& & (0.867) & (0.029) & (31.013) & (0.228) & (6.812) & (0.078) & (33.112) & (0.070) & (0.934) & (0.028) & (21.802) & (0.036) \\
\midrule
250 & 0.1 & 2.895 & 0.908 & 48.109 & 0.824 & 2.800 & 0.913 & 47.524 & 0.840 & 2.804 & 0.912 & 47.584 & 0.839 \\ 
& & (0.789) & (0.028) & (170.353) & (0.223) & (0.727) & (0.025) & (171.096) & (0.225) & (0.713) & (0.025) & (171.240) & (0.225) \\
\cmidrule{2-14}
 & 0.5 & 2.970 & 0.905 & 5.143 & 0.870 & 3.005 & 0.904 & 3.561 & 0.903 & 2.886 & 0.909 & 3.225 & 0.907 \\ 
 & & (0.519) & (0.016) & (4.321) & (0.114) & (0.447) & (0.014) & (0.740) & (0.020) & (0.451) & (0.014) & (0.629) & (0.021) \\
 \cmidrule{2-14}
  & 0.9 & 3.353 & 0.902 & 32.317 & 0.863 & 4.598 & 0.871 & 28.062 & 0.862 & 3.228 & 0.906 & 14.358 & 0.894 \\ 
 & & (1.027) & (0.028) & (29.493) & (0.080) & (1.540) & (0.032) & (27.298) & (0.030) & (1.018) & (0.028) & (15.238) & (0.024) \\
\midrule
500 & 0.1 & 3.013 & 0.901 & 3.923 & 0.882 & 2.968 & 0.905 & 3.073 & 0.906 & 2.951 & 0.905 & 3.045 & 0.907 \\ 
& & (0.751) & (0.024) & (2.718) & (0.080) & (0.666) & (0.022) & (0.654) & (0.019) & (0.701) & (0.022) & (0.672) & (0.019) \\
\cmidrule{2-14}
& 0.5 & 3.095 & 0.903 & 4.522 & 0.883 & 3.105 & 0.905 & 4.116 & 0.882 & 2.913 & 0.910 & 3.703 & 0.889 \\ 
& & (0.605) & (0.019) & (2.133) & (0.051) & (0.636) & (0.020) & (1.653) & (0.049) & (0.635) & (0.020) & (1.641) & (0.049) \\
\cmidrule{2-14}
& 0.9 & 4.136 & 0.882 & 25.779 & 0.839 & 4.741 & 0.867 & 24.110 & 0.878 & 4.041 & 0.887 & 11.840 & 0.911 \\ 
& & (3.489) & (0.073) & (31.158) & (0.238) & (3.383) & (0.068) & (30.980) & (0.020) & (3.562) & (0.074) & (18.132) & (0.016) \\
\bottomrule
\end{longtable}
\end{center}
\end{small}

\vspace{-.3in}

When the errors follow an asymmetric exponential distribution, i.e., Table~\ref{tab:tab_3}, the performance gap between the models becomes even more pronounced. Exponentially distributed errors induce skewness in the response variable, making it particularly challenging for conventional models that rely on symmetric Gaussian assumptions. As expected, ML performs the worst, with a drastic rise in MSPE and a significant drop in $R^2_{\text{test}}$ as $\rho$ increases. The FDNN model, while slightly more robust, also exhibits a notable decline in predictive performance, particularly under strong spatial dependence ($\rho = 0.9$).  In contrast, SFDNN consistently achieves the lowest MSPE across all spatial correlation levels, demonstrating its ability to adapt to asymmetric noise structures.

\begin{small}
\begin{center}
\tabcolsep 0.045in
\begin{longtable}{@{}lccccccccccccc@{}} 
\caption{\small{Comparison of ML, FDNN, and SFDNN based on MSE, MPSE, and $R^2$ metrics computed from 250 Monte Carlo simulations where the error distribution follows $\epsilon \sim \exp(1)$. Standard errors are in parentheses. Performance metrics are computed under three different training sample sizes ($n_{\text{train}}$) and three different spatial autocorrelation parameter ($\rho$)}.}\label{tab:tab_3} \\
\toprule
  &  & \multicolumn{4}{c}{ML} & \multicolumn{4}{c}{FDNN} & \multicolumn{4}{c}{SFDNN} \\
\midrule
 $n_{\text{train}}$  & $\rho$ & MSE & $R^2$ & MPSE & $R^2_{\text{test}}$ & MSE & $R^2$ & MPSE & $R^2_{\text{test}}$ & MSE & $R^2$ & MPSE & $R^2_{\text{test}}$ \\
\midrule
\endfirsthead
\toprule
  &  & \multicolumn{4}{c}{ML} & \multicolumn{4}{c}{FDNN} & \multicolumn{4}{c}{SFDNN} \\
 $n_{\text{train}}$ &  $\rho$ & MSE & $R^2$ & MPSE & $R^2_{\text{test}}$ & MSE & $R^2$ & MPSE & $R^2_{\text{test}}$ & MSE & $R^2$ & MPSE & $R^2_{\text{test}}$ \\
\midrule
\endhead
\endfoot
\endlastfoot
 100 & 0.1 & 1.002 & 0.964 & 2.500 & 0.928 & 1.006 & 0.964 & 1.350 & 0.957 & 1.010 & 0.964 & 1.357 & 0.957 \\ 
 & & (0.241) & (0.009) & (1.939) & (0.055) & (0.162) & (0.008) & (0.236) & (0.008) & (0.182) & (0.008) & (0.215) & (0.008) \\
 \cmidrule{2-14} 
  & 0.5 & 1.034 & 0.965 & 8.861 & 0.805 & 1.315 & 0.957 & 1.828 & 0.945 & 1.020 & 0.966 & 1.545 & 0.950 \\ 
 & & (0.296) & (0.009) & (15.520) & (0.311) & (0.547) & (0.013) & (0.316) & (0.347) & (0.010) & (0.282) & (0.009) & (0.009) \\
 \cmidrule{2-14}
  & 0.9 & 1.161 & 0.961 & 37.082 & 0.886 & 3.499 & 0.896 & 39.827 & 0.871 & 2.414 & 0.926 & 15.459 & 0.903 \\ 
 & & (0.322) & (0.013) & (32.332) & (0.238) & (2.771) & (0.074) & (34.476) & (0.082) & (1.933) & (0.064) & (12.341) & (0.071) \\
\midrule
250 & 0.1 & 1.129 & 0.960 & 4.668 & 0.884 & 1.024 & 0.965 & 1.263 & 0.959 & 1.019 & 0.965 & 1.248 & 0.959 \\ 
 & & (0.266) & (0.011) & (11.165) & (0.236) & (0.194) & (0.008) & (0.198) & (0.004) & (0.209) & (0.009) & (0.198) & (0.005) \\
 \cmidrule{2-14}
  & 0.5 & 1.093 & 0.963 & 2.690 & 0.934 & 1.264 & 0.958 & 1.734 & 0.952 & 1.055 & 0.965 & 1.366 & 0.956 \\ 
 & & (0.238) & (0.008) & (1.421) & (0.043) & (0.264) & (0.007) & (0.435) & (0.008) & (0.181) & (0.006) & (0.248) & (0.007) \\
 \cmidrule{2-14}
  & 0.9 & 1.062 & 0.969 & 21.107 & 0.874 & 2.937 & 0.918 & 19.973 & 0.919 & 1.172 & 0.966 & 5.464 & 0.944 \\ 
 & & (0.160) & (0.004) & (26.802) & (0.240) & (1.144) & (0.025) & (25.639) & (0.013) & (0.212) & (0.007) & (5.032) & (0.014) \\
\midrule
500 & 0.1 & 1.148 & 0.961 & 4.862 & 0.883 & 1.042 & 0.967 & 1.402 & 0.957 & 1.034 & 0.966 & 1.372 & 0.957 \\ 
 & & (0.174) & (0.006) & (11.494) & (0.238) & (0.062) & (0.003) & (0.240) & (0.008) & (0.088) & (0.003) & (0.213) & (0.008) \\
 \cmidrule{2-14}
  & 0.5 & 1.055 & 0.964 & 2.542 & 0.929 & 1.152 & 0.962 & 1.613 & 0.954 & 1.005 & 0.967 & 1.291 & 0.960 \\ 
 & & (0.163) & (0.007) & (1.663) & (0.051) & (0.157) & (0.006) & (0.496) & (0.010) & (0.177) & (0.006) & (0.247) & (0.008) \\
 \cmidrule{2-14}
  & 0.9 & 1.200 & 0.961 & 14.194 & 0.911 & 1.999 & 0.937 & 22.077 & 0.931 & 1.209 & 0.961 & 4.291 & 0.946 \\ 
 & & (0.203) & (0.006) & (22.944) & (0.072) & (0.431) & (0.010) & (24.368) & (0.010) & (0.143) & (0.005) & (4.650) & (0.020) \\
\bottomrule
\end{longtable}
\end{center}
\end{small}

\vspace{-.3in}

Across all experimental conditions, the results confirm that SFDNN significantly outperforms both ML and FDNN models in terms of predictive accuracy, particularly in the presence of high spatial dependence and non-Gaussian errors. The advantage of SFDNN is most evident when the data exhibit strong spatial autocorrelation ($\rho = 0.9$) and heavy-tailed or skewed error distributions, where conventional methods fail to provide reliable predictions.  

Some of the standard errors reported in Tables~\ref{tab:tab_1}-\ref{tab:tab_3} are relatively large, which may be attributed to data heterogeneity, strong spatial correlation, or the nonlinear complexity of the model. This is particularly pronounced in regions with sparse observations or extreme covariate values. While the model captures the general spatial-functional relationship effectively, local instability in the data may contribute to estimation variability. To reduce estimation variability, regularization techniques (e.g., weight decay, dropout), model ensembling, or spatial smoothing priors can be incorporated into the proposed approach.

\section{Application to Brazilian COVID-19 data}\label{sec6}

Understanding the spatial and temporal dynamics of infectious diseases is crucial for effective public health interventions. In this section, we analyze a large-scale dataset comprising COVID-19 statistics from 5,570 Brazilian cities, providing a detailed examination of how geographic proximity and nonlinear relationships influence pandemic outcomes. The dataset, obtained through the \texttt{COVID19} \Rlogo\ package \citep{COVID19}, includes daily death counts per city as the scalar response variable, while three functional predictors—daily confirmed cases, daily people vaccinated, and daily people fully vaccinated—capture the evolving nature of the pandemic. Additionally, we incorporate city population as a scalar predictor to account for the varying scales of COVID-19 impact across urban and rural regions. The dataset spans the years 2021 and 2022, enabling us to analyze trends over time and assess the long-term effects of vaccination efforts and public health measures.

Traditional regression models often assume independence across observations, failing to capture the extent to which a city's outbreak trajectory is influenced by its neighboring areas. Given the high population density and extensive intercity travel in Brazil, ignoring spatial autocorrelation can lead to misleading conclusions about the effectiveness of containment measures, vaccination campaigns, and policy decisions. Incorporating spatial correlation allows us to better estimate the true impact of interventions, as adjacent cities are more likely to experience similar epidemic trends due to shared resources, healthcare accessibility, and human mobility patterns \citep[see, e.g.,][]{BHGARC2024, BSM2025}.

\begin{figure}[!htb]
\centering
\includegraphics[width=9cm]{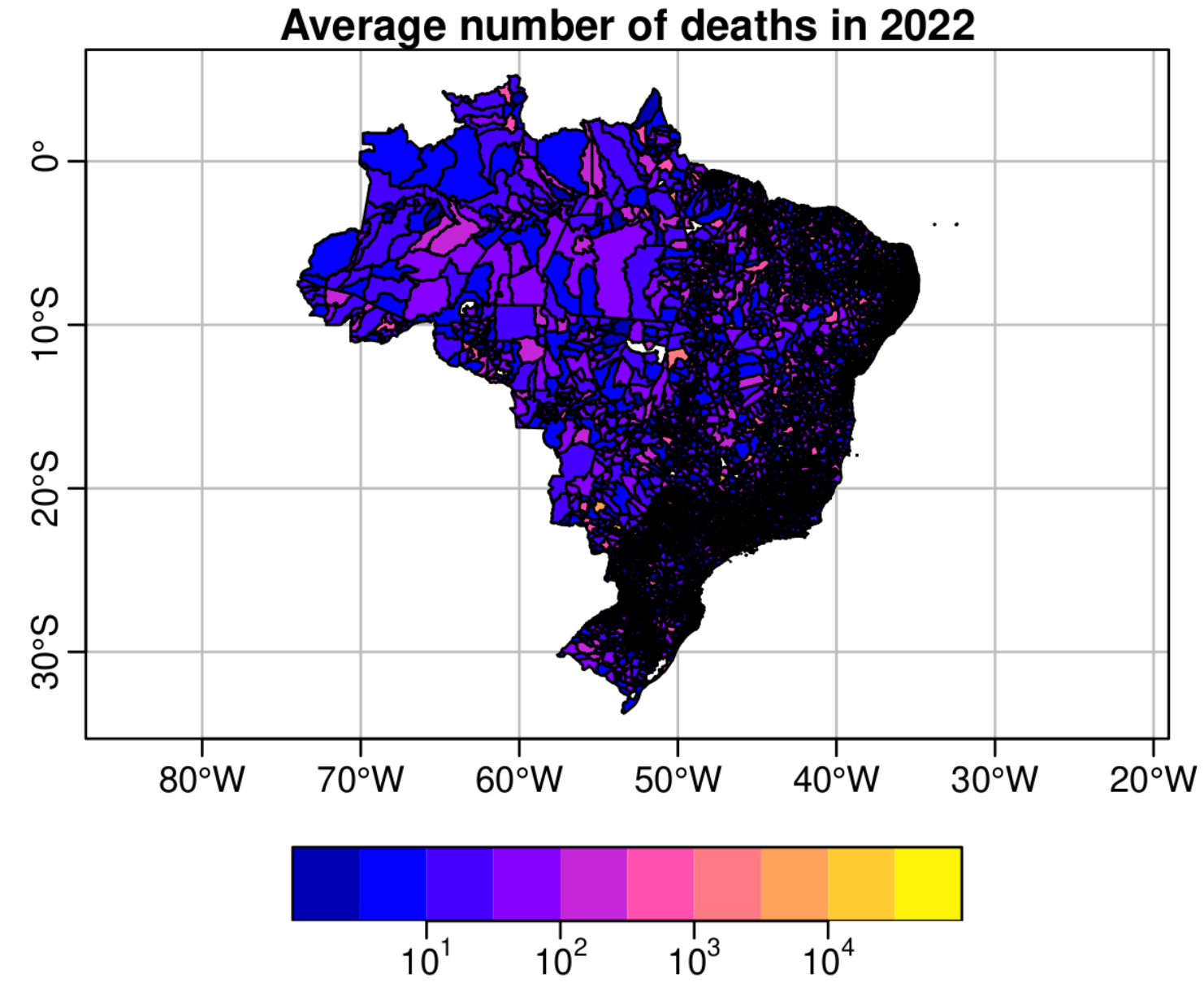}
\quad
\includegraphics[width=9cm]{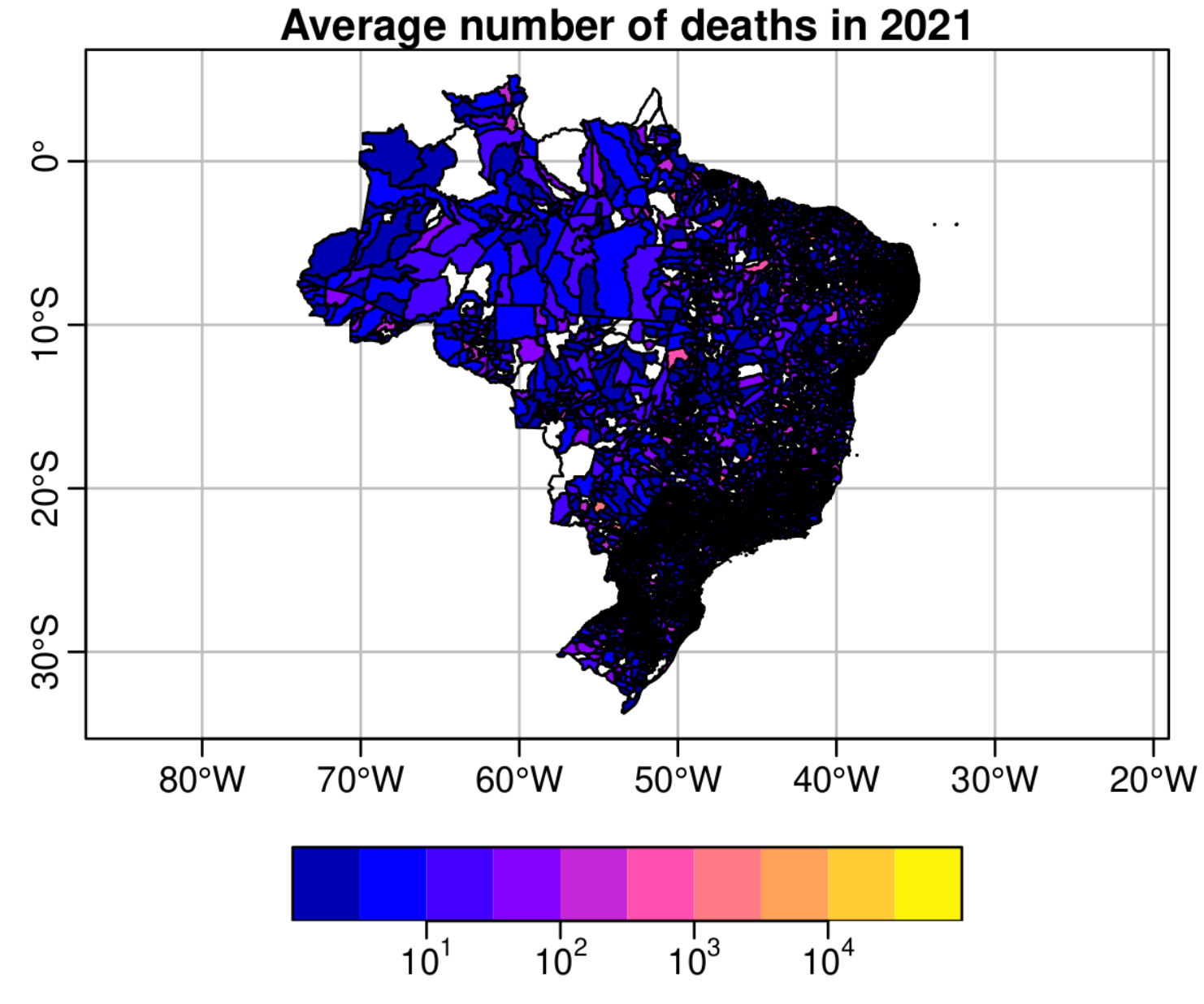}
\caption{\small{Choropleth maps depicting the average number of
deaths for Brazilian cities in the years 2021 (left panel) and 2022 (right panel)}.}
\label{fig:Fig_3}
\end{figure}

In Figure~\ref{fig:Fig_3}, we present choropleth maps illustrating the spatial distribution of COVID-19 deaths across Brazilian cities for the years 2021 and 2022. The maps reveal a distinct clustering pattern, indicating that mortality rates are not randomly distributed but rather exhibit strong spatial dependence. Specifically, cities with high death counts tend to be geographically close to other high-mortality regions, suggesting that the spread and severity of the virus are significantly influenced by local transmission dynamics, healthcare infrastructure, and regional policy measures. This spatial clustering underscores the necessity of explicitly modeling spatial correlation in mortality data.

A critical limitation of classical models is their assumption of linear relationships between predictors and outcomes. However, in the case of COVID-19, both the interaction between predictors and their individual effects on daily death counts may be highly nonlinear. The number of confirmed cases and vaccination rates, for instance, do not exert a simple additive influence on mortality. Instead, their combined effect may exhibit threshold behaviors, saturation effects, or synergy, as suggested by \cite{BSM2024}, which found strong interaction effects between vaccination rates and confirmed case counts in COVID-19 data.

\begin{figure}[!htb]
\centering
\includegraphics[width=5.83cm]{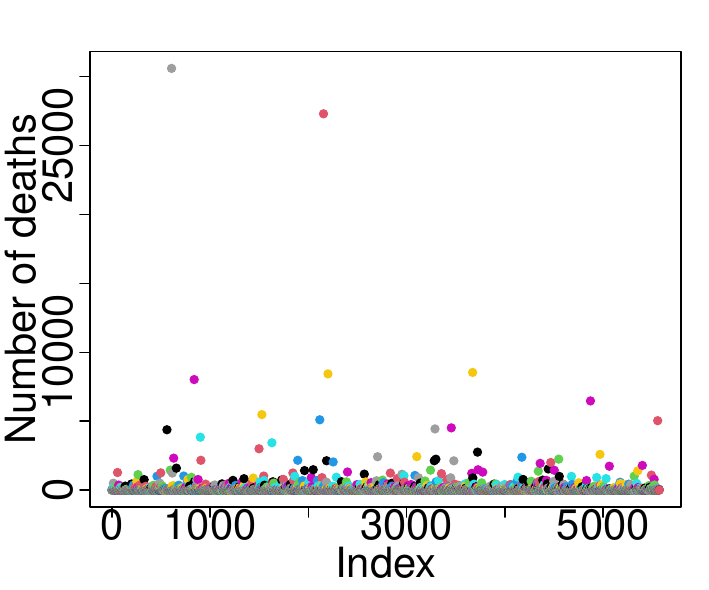}
\quad
\includegraphics[width=5.83cm]{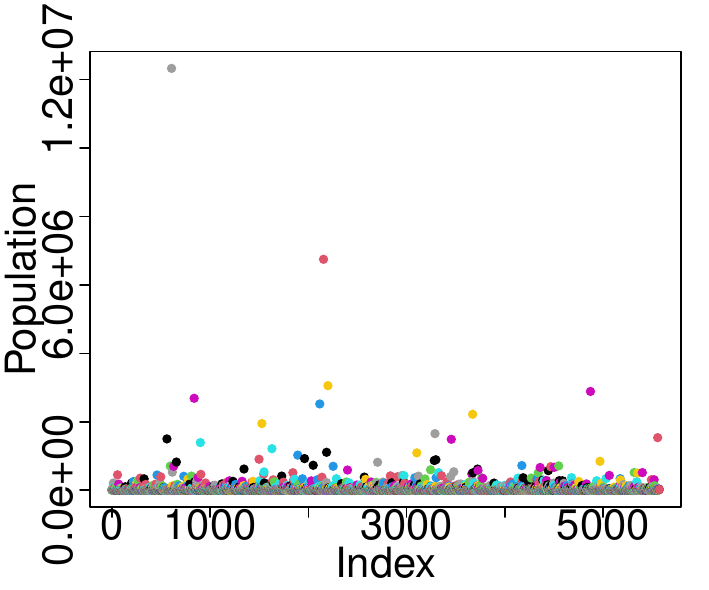}
\\
\includegraphics[width=5.83cm]{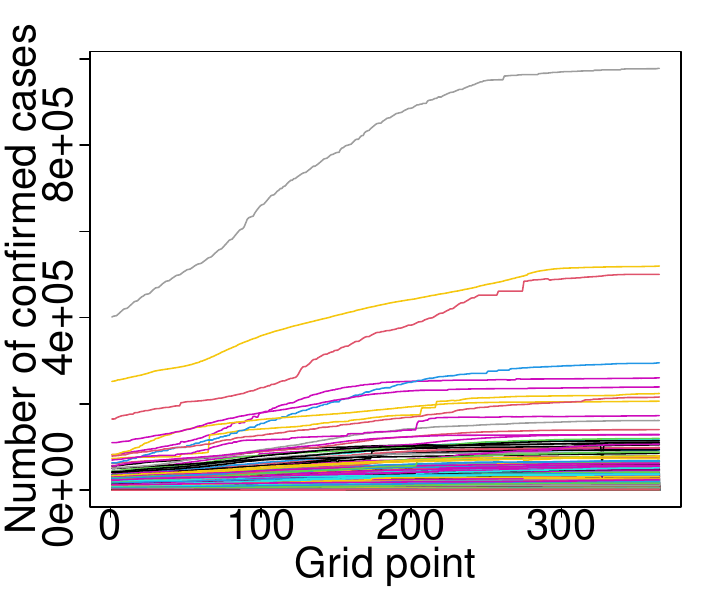}
\quad
\includegraphics[width=5.83cm]{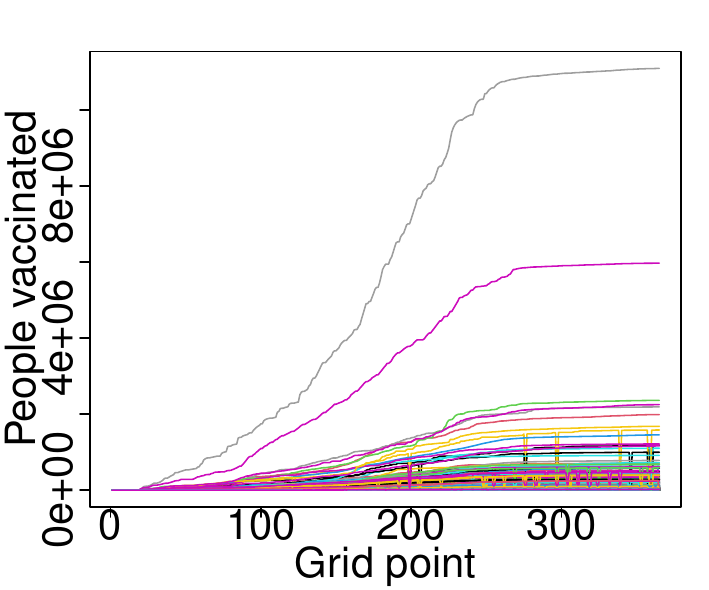}
\quad
\includegraphics[width=5.83cm]{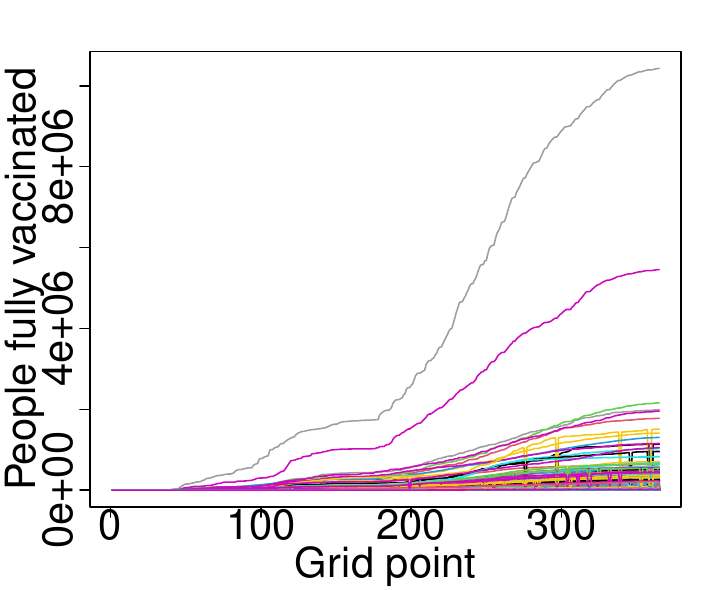}
\caption{\small{The graphical display of Brazil's COVID-19 data presents the confirmed cases (top left panel), population (top right panel), number of confirmed cases (bottom left panel), people vaccinated (bottom middle panel), and people fully vaccinated (bottom right panel) for 2021. Different colors represent different cities, with the observations of confirmed cases, people vaccinated, and people fully vaccinated being functions of days, i.e., $1 \leq t \leq 365$}.}\label{fig:Fig_4}
\end{figure}

By incorporating spatial correlation and nonlinear interactions into our modeling framework, we provide a more realistic prediction mechanism for COVID-19 mortality in Brazil. The proposed SSFDNN allows us to flexibly model these complex dependencies, capturing high-dimensional relationships that standard linear models fail to identify. Figure~\ref{fig:Fig_4} provides a graphical representation of the variables, revealing potential outliers in both the response and predictor variables. To mitigate the effects of extreme values and improve modeling stability, we applied a natural logarithm transformation to all scalar and functional variables in the COVID-19 dataset. This transformation is commonly used when variables exhibit positive skewness or heavy right tails, as it stabilizes variance and reduces the influence of outliers. We also considered alternative transformations, such as the Box-Cox family, which generalizes the log transform by allowing for a continuum of power transformations. However, empirical evaluation using cross-validated predictive performance showed negligible improvements compared to the log-transform, while the latter retained interpretability and simplicity. Given that all relevant variables (e.g., death counts, case counts, vaccinations, population) are strictly positive and right-skewed, the log transformation provided a statistically sound and computationally efficient means of stabilizing variance and enhancing model robustness.

To define the spatial weight matrix $\bm{W} = (w_{ii^\prime})_{5570 \times 5570}$ for the network of 5570 cities in Brazil, we implement a K-nearest neighbors method with a bi-square weighting scheme and an adaptive bandwidth selection. Initially, the geographic coordinates (latitude and longitude) of each city are used to calculate the pairwise great-circle distances $\{d_{ii^\prime}\}$ between any two locations $i$ and $i^\prime$. For each city $i$, $h$ closest neighbors are identified based on these distances, where $h$ is determined through cross-validation. The adaptive bandwidth $H_i$ is set as the largest distance among these four nearest neighbors, given by $H_i = \max \{ d_{ii^\prime} : i^\prime \in N_4(i) \}$, where $N_4(i)$ represents the set of city $i$'s $h$ closest neighbors. The spatial weight assigned to each neighboring city $i^\prime$ is calculated using a bi-square kernel function:
\begin{equation*}
w_{ii^\prime} = \left( 1 - \left( \frac{d_{ij}}{H_i} \right)^2 \right)^2, \quad \text{for } i^\prime \in N_h(i).    
\end{equation*}
Cities that do not fall within the $h$ nearest neighbors of $i$ are assigned a weight of zero ($w_{ii^\prime} = 0$ for $i^\prime \notin N_h(i)$). Finally, to maintain consistency in the weighting scheme, the assigned weights are normalized such that they sum to one for each city $w_{ii^\prime} = \frac{w_{ii^\prime}}{\sum_{i^\prime \in N_h(i)} w_{ii^\prime}}$.

Next, we calculate the local Moran’s I statistic \citep{Ansellin1995}, which is designed to assess the degree of significant spatial clustering of similar values around each location. For the $i$th city, the local Moran’s I statistic is formulated as follows:
\begin{equation*}
I_i = \frac{n (Y_i - \overline{Y})}{\sum_{i^{\prime} = 1}^n (Y_{i^{\prime}} - \overline{Y})^2} \sum_{i^{\prime} = 1}^n w_{i i^{\prime}} (Y_{i^{\prime}} - \overline{Y}).
\end{equation*}
In applied settings, local Moran’s I values are frequently utilized for spatial visualization, revealing regions where strong or weak spatial dependencies exist. A high $I_i$ value indicates that the given location belongs to a cluster of similar values, which may be either high, low, or moderate. In contrast, a low $I_i$ value suggests that the area is surrounded by locations with substantially different values. This classification identifies distinct spatial clusters, commonly categorized as High-High, Low-Low, High-Low, and Low-High.
\begin{figure}[!htb]
\centering
\includegraphics[width=12cm]{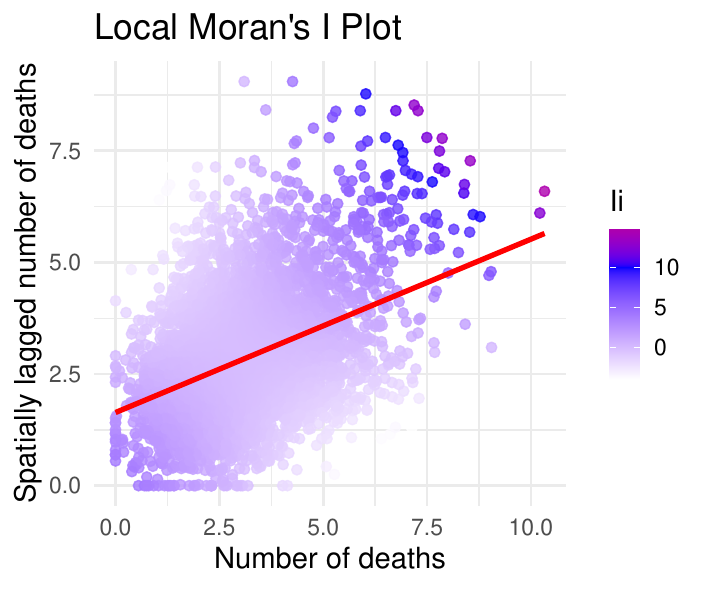}
\caption{\small{Scatter plot of Moran’s I, depicting the local spatial autocorrelation of the average number of deaths across 5570 cities in Brazil}.}
\label{fig:Fig_5}
\end{figure}

The obtained local Moran’s I values are illustrated in Figure~\ref{fig:Fig_5}. The positive slope observed in the regression line within this plot indicates an overall presence of positive spatial autocorrelation in the data. This suggests that the number of deaths values tends to be spatially grouped, with high values clustering near other high values and low values appearing close to other low values.

The ML, FDNN, and the proposed SFDNN models are applied to the Brazilian COVID-19 dataset for the year 2021 to train the models. Subsequently, the fitted models utilize scalar and functional predictors from the year 2022 to forecast the average number of deaths for 2024. Our findings reveal that all three ML-based methods yield comparable MSE values during the training phase, with ML, FDNN, and SFDNN achieving MSE values of 0.192, 0.184, and 0.168, respectively. However, in terms of predictive accuracy, the proposed SFDNN model outperforms the alternatives, achieving the lowest MSPE of 0.253. In contrast, the ML and FDNN models exhibit higher RMSE values of 0.708 and 1.106, respectively. These results underscore the superior performance of the SFDNN model in effectively capturing complex spatial and functional dependencies in the Brazilian COVID-19 dataset.

\begin{figure}[!htb]
\centering
\includegraphics[width=9cm]{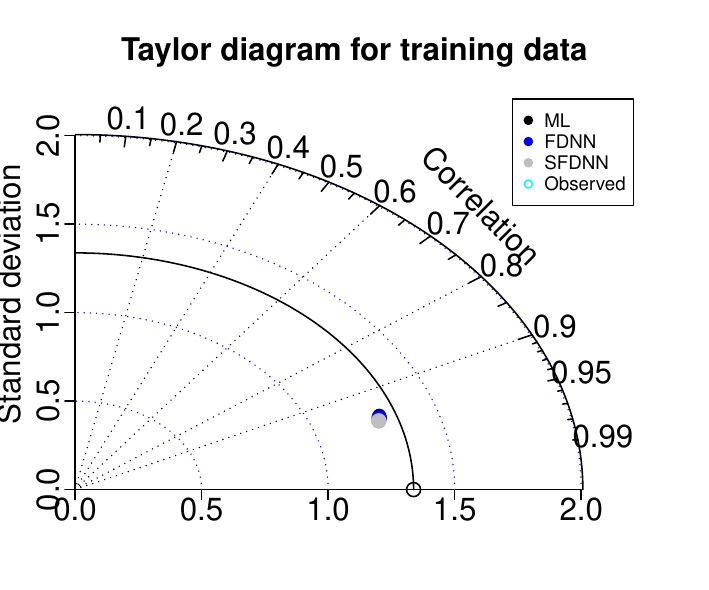}
\quad
\includegraphics[width=9cm]{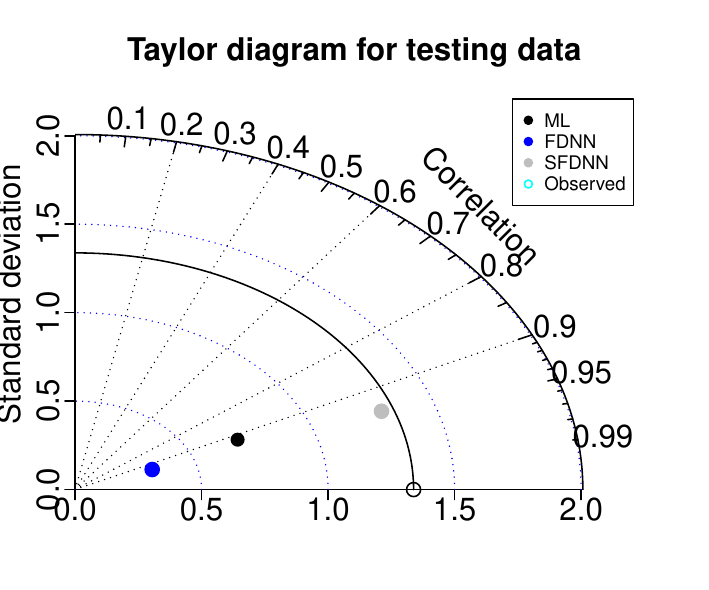}
\caption{\small{Taylor diagrams comparing the performance of the ML, FDNN, and SFDNN models for training (left) and testing (right) datasets in predicting the number of deaths in Brazil}.}
\label{fig:Fig_6}
\end{figure}

Taylor diagrams in Figure~\ref{fig:Fig_6} provide an insightful evaluation of model performance in both training and testing phases, where SFDNN exhibits a stronger alignment with the observed data, as reflected by its higher correlation and improved variance representation. While ML and FDNN models also capture spatial dependencies to some extent, their correlation with observed values remains slightly lower, indicating that they are less effective in modeling the intricate spatial-functional interactions present in the dataset.

\begin{figure}[!htb]
\centering
\includegraphics[width=5.8cm]{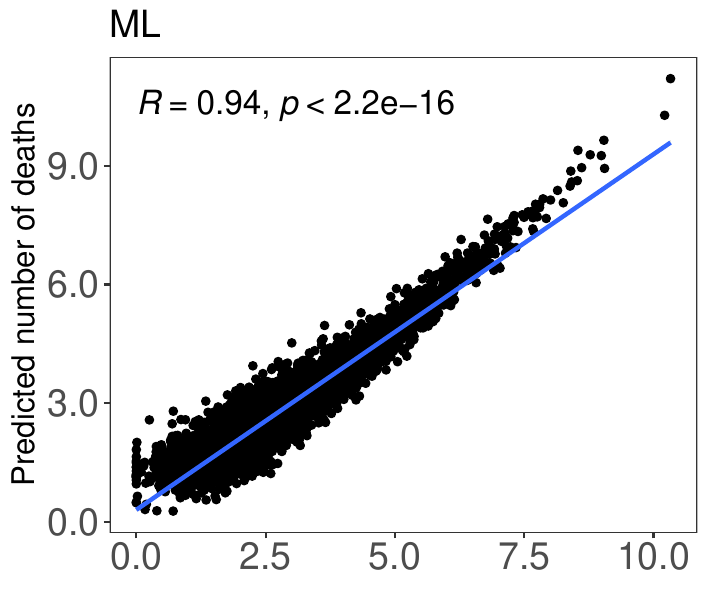}
\quad
\includegraphics[width=5.8cm]{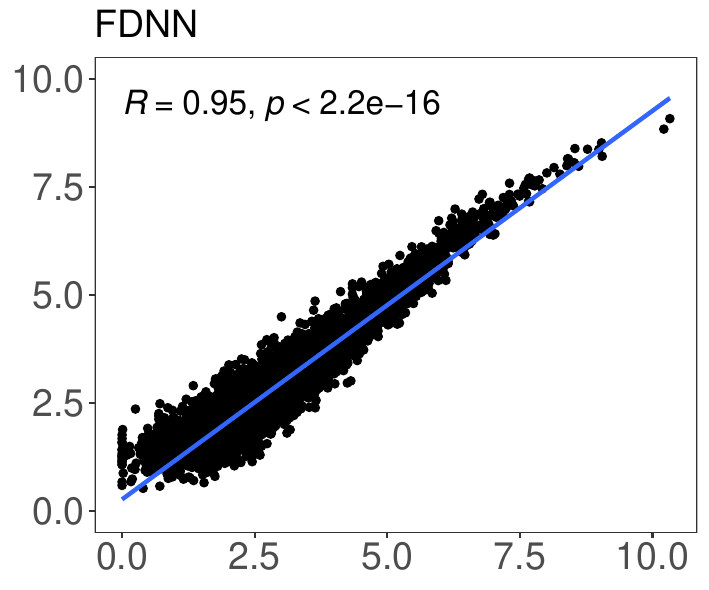}
\quad
\includegraphics[width=5.8cm]{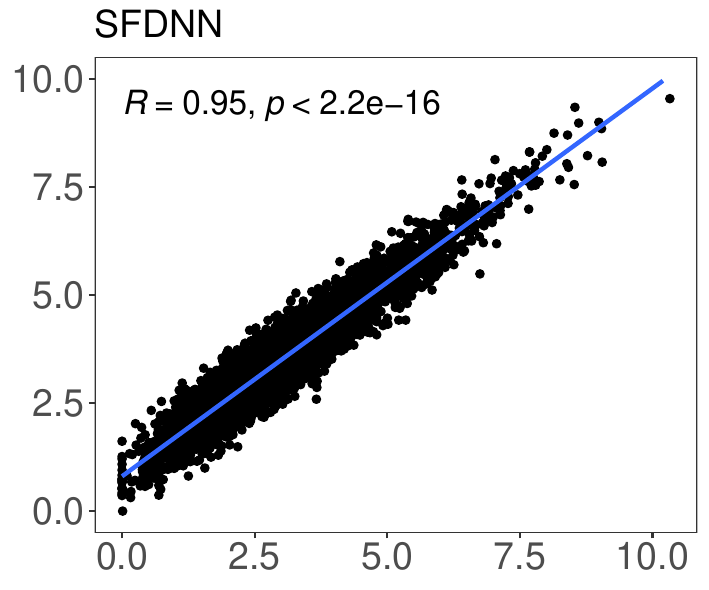}
\\  
\includegraphics[width=5.8cm]{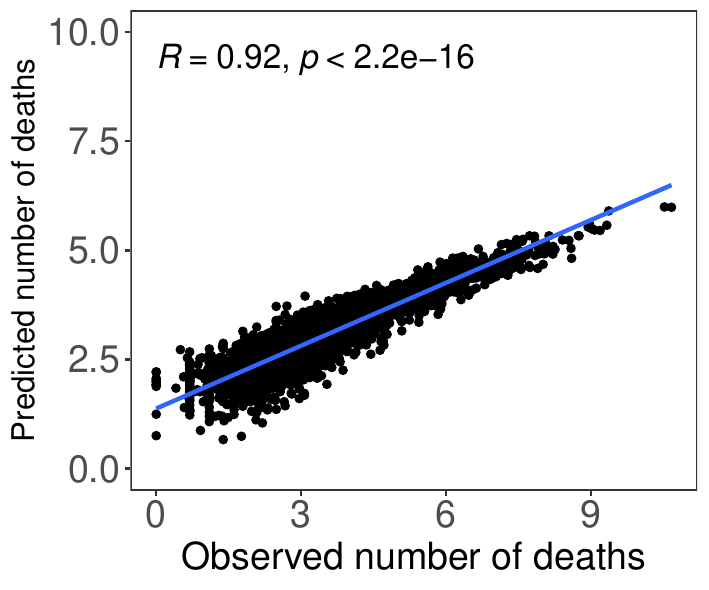}
\quad
\includegraphics[width=5.8cm]{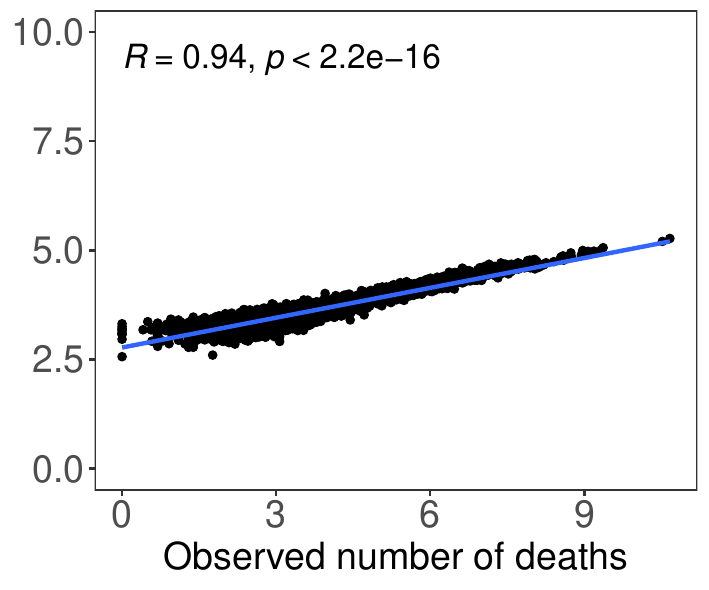}
\quad
\includegraphics[width=5.8cm]{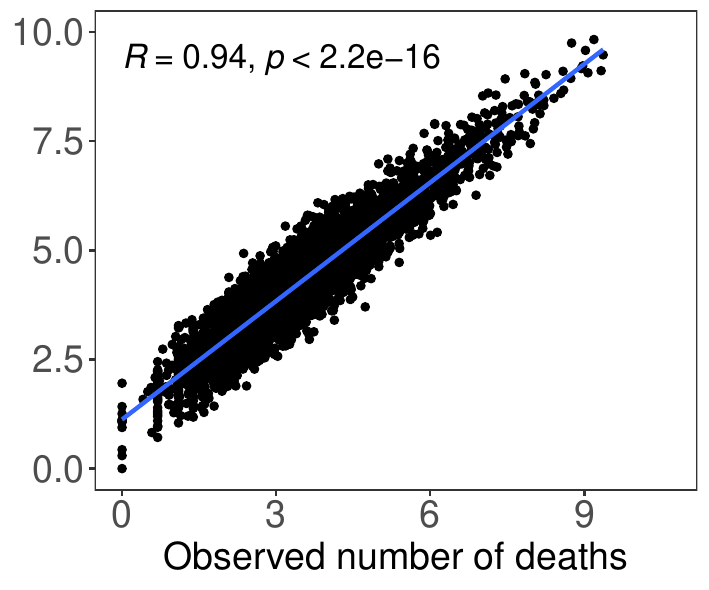}
\caption{\small{Scatter plots of observed vs. predicted number of deaths values for the Brazilian COVID-19 dataset. The first row represents model performance on the training dataset (2021), while the second row shows results for the testing period (2022). The three columns correspond to the ML, FDNN, and SFDNN models, respectively. Each plot includes a fitted regression line, and Pearson correlation coefficients ($R$) indicate the strength of association between observed and predicted values}.}\label{fig:Fig_7}
\end{figure}

In Figure~\ref{fig:Fig_7}, further evidence of the SFDNN model's predictive advantage is observed in the scatter plots of observed vs. predicted values. From this plot, while all three methods exhibit high correlations between observed and predicted values, the SFDNN model stands out for its superior predictive accuracy and tighter clustering of points around the fitted regression line. This improvement arises from SFDNN’s ability to capture both nonlinear interaction effects among the functional predictors and the spatial dependencies inherent in the data—factors that standard ML and FDNN approaches address less effectively. The elevated correlation coefficients (up to $R = 0.95$) underscore SFDNN’s robust performance, particularly for testing data, suggesting that modeling spatial and functional interactions is crucial for reliable mortality forecasting in large-scale epidemiological applications.

\section{Conclusion}\label{sec7}

In this study, a novel nonlinear modeling framework is designed to effectively capture both spatial dependencies and functional relationships in high-dimensional data. By integrating a spatial autoregressive structure with a deep learning-based functional regression approach, the SFDNN model provides a flexible and robust solution for predictive modeling in spatiotemporal domains. Unlike conventional scalar-on-function regression models that assume linearity and independence across observations, SFDNN accommodates complex nonlinear dependencies and spatial correlations, significantly enhancing predictive accuracy.

Through extensive Monte Carlo simulations, the SFDNN model outperformed the existing ML and FDNN models, particularly in scenarios with strong spatial dependencies. The results showed that while all models performed comparably in training, SFDNN exhibited the lowest MSPE in the testing phase, highlighting its superior generalization capability. The empirical application to Brazilian COVID-19 data further validated the findings, where SFDNN achieved the highest predictive accuracy for the average number of deaths. 

The contributions of this study extend beyond prediction performance, offering key insights into spatial-functional learning and its broader applications that include: i) By incorporating a spatial dependence structure into a deep neural network, the proposed model effectively learns both spatial and functional interactions, making it particularly valuable for spatiotemporal forecasting; ii) The SFDNN model addresses a fundamental limitation of classical regression models by allowing for flexible nonlinear transformations, crucial for capturing real-world phenomena; iii) The model’s ability to handle multiple functional and scalar predictors in a data-driven manner makes it a scalable and generalizable tool for various domains, including population health, geosciences, climate modeling, and environmental monitoring.

Despite its strengths, several areas warrant further exploration: i) Training deep neural networks with spatially dependent functional data requires more computational resources (compared with traditional methods). Future research could explore efficient optimization techniques and GPU-accelerated training to enhance scalability; ii) The current study utilizes a K-nearest neighbors bi-square kernel for spatial weight construction, future studies may explore adaptive spatial kernels or data-driven approaches to further refine spatial dependence modeling; While this study focuses on spatial functional data, an important extension would be to incorporate time-varying dependencies leading to a fully spatiotemporal functional deep learning model; iii) Although deep learning models provide powerful predictive capabilities, incorporating Bayesian deep learning or interval prediction methods could offer additional insights into the uncertainty of predictions; iv) Although the proposed SFDNN model involves computationally intensive operations—such as inverting the spatial matrix $(\mathbb{I}_n - \rho \bm{W})$ and backpropagating through functional layers—these can be efficiently optimized using sparse linear algebra techniques (e.g., Cholesky or LU decomposition for sparse matrices) and parallelized via GPU acceleration to accommodate large-scale datasets.

\section*{Conflicts of interests/Competing interests}

The authors have no conflicts of interest to declare that are relevant to the content of this article.

\section*{Data Availability}

The data used in this study are publicly available through the \texttt{COVID19} \Rlogo\ package, which provides access to official COVID-19 datasets from multiple sources. The package and its documentation are available at \url{https://CRAN.R-project.org/package=COVID19}.

\section*{Acknowledgments}

This research was supported by the Scientific and Technological Research Council of Turkey (TUBITAK) (grant no. 124F096) and the Australian Research Council Future Fellowship (grant no. FT240100338). 

\bibliographystyle{agsm}
\bibliography{sfof.bib}

\end{document}